%% file: template.tex
\documentclass[10pt,journal,cspaper,compsoc]{IEEEtran}
\pdfoutput=1

\ifCLASSOPTIONcompsoc
  \usepackage[nocompress]{cite}
\else
  \usepackage{cite}
\fi
\ifCLASSINFOpdf
   \usepackage[pdftex]{graphicx}
   \graphicspath{{./pictures/}{./pictures/encoding/}{./pictures/results/}{./pictures/evaluation/}{./pictures/biography/}}
   \DeclareGraphicsExtensions{.pdf,.jpeg,.jpg,.png,.eps}
\else
\fi

\usepackage{hyperref}

\usepackage{subcaption}
\usepackage{wrapfig}
\usepackage{amsmath}
\usepackage{algorithm}
\usepackage[noend]{algpseudocode}

\usepackage[percent]{overpic}

\usepackage[dvipsnames]{xcolor}
\usepackage{multirow}

\usepackage{microtype}                 
\PassOptionsToPackage{warn}{textcomp}  
\usepackage{textcomp}                  
\usepackage{mathptmx}                  
\usepackage{times}                     
\usepackage{cite}                      
\usepackage{tabu}                      
\usepackage{booktabs}                  
\newcommand{\stream}{\textit{NSG}}
\newcommand{\tree}{\textit{Treemaps}}
\newcommand{\secs}{\textit{SplitStreams}}
\newcommand{\tme}{\textit{total time per task}}
\newcommand{\err}{\textit{error of first response}}
\newcommand{\attempts}{\textit{number of attempts}}
\newcommand{\attemptTime}{\textit{time for first response}}

\begin{document}
%
\title{SplitStreams: A Visual\\Metaphor for Evolving Hierarchies}
%
%
%
%

\author{Fabian~Bolte,~\IEEEmembership{Member,~IEEE,}
        Mahsan~Nourani,~\IEEEmembership{Member,~IEEE,}\\
        Eric~D.~Ragan,~\IEEEmembership{Member,~IEEE,}
        and~Stefan~Bruckner,~\IEEEmembership{Member,~IEEE}
\IEEEcompsocitemizethanks{\IEEEcompsocthanksitem Fabian~Bolte and Stefan~Bruckner are with the Department of Informatics, University of Bergen, Norway.\protect\\
E-mail: \{fabian.bolte, stefan.bruckner\}@uib.no
}
\IEEEcompsocitemizethanks{\IEEEcompsocthanksitem Mahsan~Nourani and Eric~D.~Ragan are with the Department of Computer \& Information Science \& Engineering, University of Florida, United States.\protect\\
E-mail: \{mahsannourani, eragan\}@ufl.edu
}
}

%
%

\markboth{Journal of \LaTeX\ Class Files,~Vol.~14, No.~8, August~2015}%
{Shell \MakeLowercase{\textit{et al.}}: Bare Demo of IEEEtran.cls for Computer Society Journals}
%



\IEEEtitleabstractindextext{%
\begin{abstract}
The visualization of hierarchically structured data over time is an ongoing challenge and several approaches exist trying to solve it. Techniques such as animated or juxtaposed tree visualizations are not capable of providing a good overview of the time series and lack expressiveness in conveying changes over time. Nested streamgraphs provide a better understanding of the data evolution, but lack the clear outline of hierarchical structures at a given timestep. Furthermore, these approaches are often limited to static hierarchies or exclude complex hierarchical changes in the data, limiting their use cases. We propose a novel visual metaphor capable of providing a static overview of all hierarchical changes over time, as well as clearly outlining the hierarchical structure at each individual time step. Our method allows for smooth transitions between tree maps and nested streamgraphs, enabling the exploration of the trade-off between dynamic behavior and hierarchical structure. As our technique handles topological changes of all types, it is suitable for a wide range of applications. We demonstrate the utility of our method on several use cases, evaluate it with a user study, and provide its full source code.%
\end{abstract}

}

\maketitle

\IEEEdisplaynontitleabstractindextext

%
\IEEEpeerreviewmaketitle

\IEEEraisesectionheading{\section{Introduction}\label{sec:introduction}}

%
%
%
%
\IEEEPARstart{H}{ierarchically}
\input{s1_introduction.tex}
\input{s2_relatedWork.tex}
\input{s3_overview.tex}
\input{s4_visualDesign.tex}
\input{s6_results.tex}

\input{s7_evaluation.tex}
\input{s8_discussion.tex}
\input{s9_conclusion.tex}

\appendices
\label{appendix}
\section{}
In the following we describe \autoref{alg:generate} from \autoref{sec:visualDesign}.4 in pseudocode.
\begin{algorithm}
\caption{Streamgraph Generation}\label{alg:generate}
    \begin{algorithmic}[1]
        \Procedure{traverseStream}{node}
            \If{\textit{All previous nodes of node have been visited}}
                \State \textit{mark node as visited}
            \EndIf
            \If{\textit{Node has next node}}
                \ForAll{\textit{next}}
                    \If{\textbf{not} \textit{next was visited}}
                        \State $t1 \gets t(node) + 0.5 \times HCR \times (t(next) - t(node))$
                        \State $t2 \gets t(next) - 0.5 \times HCR \times (t(next) - t(node))$
                        \State \textit{draw straight stream from t(node) to t1}
                        \State \textit{draw curved stream from t1 to t2}
                        \State \textit{draw straight stream from t2 to t(next)}
                        \State \Call{traverseStream}{next}
                    \EndIf
                \EndFor
            \Else
                \State $tEnd \gets t(node) + 0.5 \times HCR$
                \State \textit{draw end cap from t(node) to tEnd}
            \EndIf
        \EndProcedure
        \State

        \Procedure{drawStreams}{}
            \ForAll{\textit{streams}}
                \State $tStart \gets t(firstNode) - 0.5 \times HCR$
                \State \textit{draw start cap from tStart to t(firstNode)}
                \State \Call{traverseStream}{\textit{root}}
            \EndFor
        \EndProcedure
    \end{algorithmic}
\end{algorithm}

\ifCLASSOPTIONcompsoc
\section*{Acknowledgments}
\else
\section*{Acknowledgment}
\fi

The research presented in this paper was supported by the MetaVis project (\#250133) funded by the Research Council of Norway.
The work is also supported in part by NSF 1565725 and the DARPA XAI program N66001-17-2-4032.

\ifCLASSOPTIONcaptionsoff
  \newpage
\fi



\bibliographystyle{IEEEtran}
\bibliography{IEEEabrv,template}
%



%

\vspace{-0.25cm}
\begin{IEEEbiography}[{\includegraphics[width=1in,height=1.25in,clip,keepaspectratio]{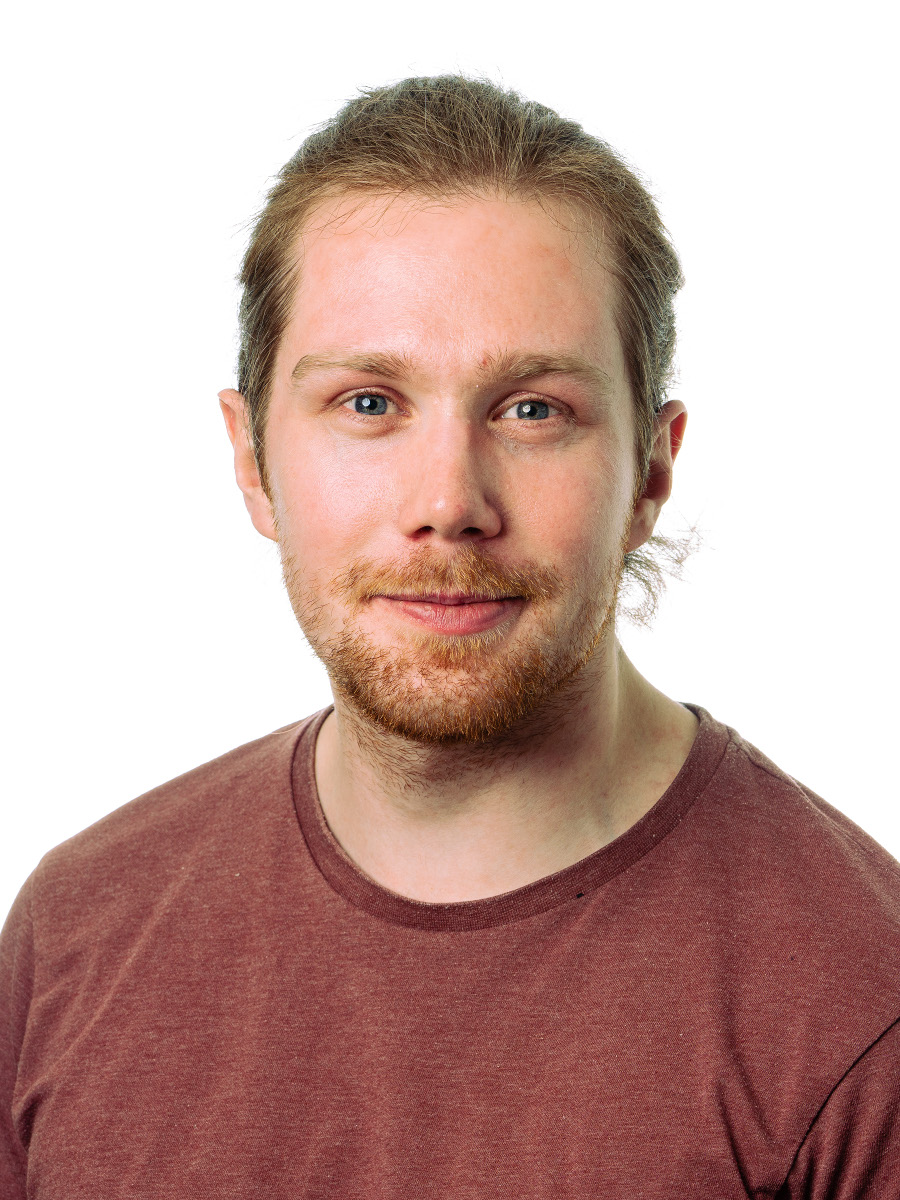}}]{Fabian Bolte}
is a PhD student in visualization at the Department of Informatics of the University of Bergen, Norway. He received his bachelor's degree in Applied Computer Science from the TU Chemnitz, Germany in 2014 and his master's degree in High Performance \& Cloud Computing in 2016 from the same university. His research interests include the visualization of time-dependent data, visual parameter space analysis, and meta visualization.
\end{IEEEbiography}

\vspace{-1cm}
\begin{IEEEbiography}[{\includegraphics[width=1in,height=1.25in,clip,keepaspectratio]{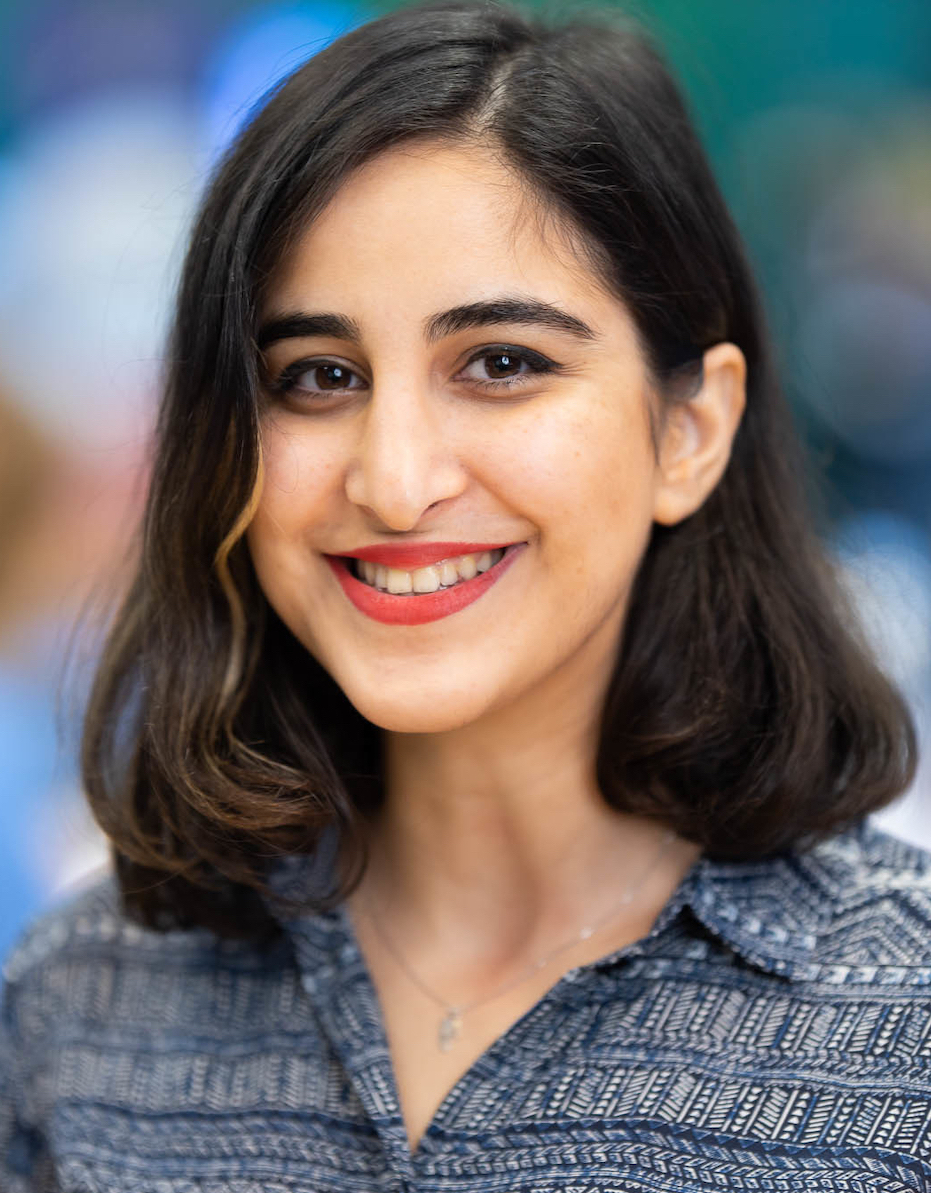}}]{Mahsan Nourani}
is a PhD student in Computer Science at the Department of Computer \& Information Science \& Engineering at the University of Florida, United States.
She received her bachelor's degree in Computer Engineering from the University of Tehran, Iran in 2017.
Her research interests include Human-Computer Interaction, user-centered design, human-centered AI, explainable intelligent systems, and visual analytics.
\end{IEEEbiography}

\vspace{-0.75cm}
\begin{IEEEbiography}[{\includegraphics[width=1in,height=1.25in,clip,keepaspectratio]{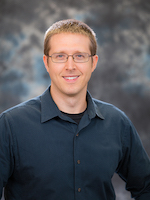}}]{Eric D. Ragan}
is an Assistant Professor in the Department of Computer \& Information Science \& Engineering at the University of Florida, United States. He directs the Interactive Data and Immersive Environments (INDIE) lab, which conducts research of human-computer interaction, visual analytics, virtual reality, and explainable intelligent systems. He received his Ph.D. in Computer Science from Virginia Tech. He is a member of the IEEE Computer Society.
\end{IEEEbiography}

\vspace{-1cm}
\begin{IEEEbiography}[{\includegraphics[width=1in,height=1.25in,clip,keepaspectratio]{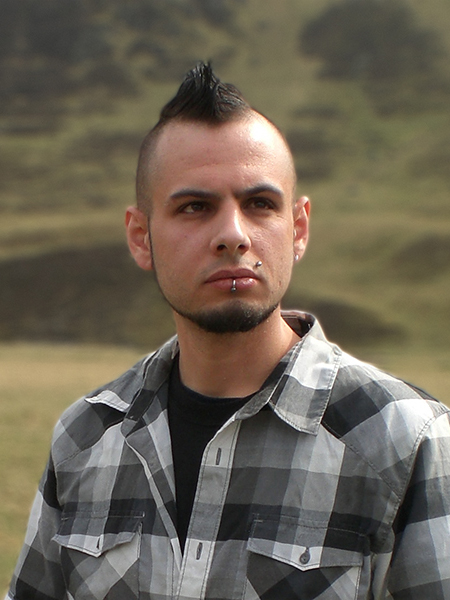}}]{Stefan Bruckner}
is a professor in visualization at the Department of Informatics of the University of Bergen, Norway. He received his master’s degree in Computer Science from the TU Wien, Austria in 2004 and his Ph.D. in 2008 from the same university. He was awarded the habilitation (venia docendi) in Practical Computer Science in 2012.
From 2008 to 2013, he was an assistant professor at the Institute of Computer Graphics and Algorithms at TU Wien. His research interests include interactive visualization techniques for spatial data, particularly in the context of biomedical applications, visual parameter space analysis, illustrative methods, volume visualization, and knowledge-assisted visual interfaces.
\end{IEEEbiography}

\vfill








\end{document}

%% file: s1_introduction.tex
 structured data can be found in many places and has been addressed by many visualization techniques. Ancestry, taxonomy, topology, company staff, file systems, text articles, source code, and population data are just a few examples of data with inherent hierarchical structures that can be represented by a tree with parent-child relationships. 
 Data that does not have inherent hierarchical structure, such as scalar data~\cite{heine2016survey} and large text corpora~\cite{cui2014hierarchical}, can be clustered to build a hierarchy for providing a better overview and understanding of the data. The corresponding trees have previously been visualized by explicit and implicit methods in two- and three-dimensional space, as well as hybrids, in all kinds of visual layouts~\cite{schulz2011treevis}.
 Visualizing the evolution of such data, its hierarchical structure, and temporal changes requires the integration of time as yet another dimension. Several approaches have utilized existing tree visualization techniques to display each timestep in either a juxtaposed layout or as an animation. 
 Further efforts have been made to optimize these layouts with respect to stability of object positions over time for preserving the users' mental map~\cite{tu2007visualizing, sud2010fast, tak2013enhanced, hahn2014visualization, van2017stable, sondag2018stable}. 
 More recent approaches began to adopt the ThemeRiver~\cite{havre2002themeriver} metaphor to convey the evolution of tree nodes in hierarchically structured data by individual streams~\cite{cuenca2018multistream, lukasczyk2017nested, kopp2019temporal, wittenhagen2016chronicler}. 
 All these methods are either limited by only supporting a static hierarchy over the whole time series~\cite{cuenca2018multistream}, only allowing for data where each parent's value is sufficiently larger than the sum of its children's values~\cite{lukasczyk2017nested}, or only allowing for a subset of trackable changes in the hierarchy~\cite{kopp2019temporal}. 
While these techniques do a good job in presenting hierarchical changes over time, they suffer from deficiencies in conveying the hierarchical structure at specific timesteps when no hierarchical changes are in view and create ambiguities in the interpretation.

We present a novel visual metaphor for representing time-dependent, hierarchically structured data in a static visualization. 
We take a nested streamgraph~\cite{lukasczyk2017nested} as a basis, introduce \textit{splits} to cut the streams at certain points in time, and add a horizontal margin. By increasing the margin based on a stream's hierarchy level, we reveal the underlying hierarchical structure of the data.
We enable fine-grained control between visual continuity of individual streams and the visual clarity of hierarchical structures at a given point in time. 
The presented approach allows for a clearer representation of hierarchies even in cases when color cannot be used to encode the hierarchy level. 
In contrast to previous approaches such as Chronicler~\cite{wittenhagen2016chronicler}, our method handles complex hierarchical changes, like a node becoming the parent of its ancestor, in a consistent and unambiguous manner. 
We propose a novel visual encoding for such cases and, in conclusion, cover all hierarchical changes that occur in the visualization of hierarchically structured data over time. 
Our main contributions can be summarized as follows:
\begin{enumerate}
  \item We introduce a novel visual metaphor to emphasize the data-inherent hierarchical structure and its changes over time. Our approach is based on simple and clear shapes that are capable of conveying the hierarchical structure independent of the color scheme applied.
  \item We conducted a user study to evaluate users' performance in analyzing hierarchically structured data with the help of treemaps over time, nested streamgraphs, and SplitStreams and demonstrate that our approach successfully addresses deficiencies of previous methods.
  \item We publish our implementation as an open source library for easy reproduction of existing visualization techniques like one-dimensional treemaps over time and nested streamgraphs, as well as the exploration of novel visualization techniques introduced by this paper.
\end{enumerate}

%% file: s2_relatedWork.tex
\section{Related Work}
\label{sec:related}

Our visual metaphor allows for the smooth transition between one-dimensional treemaps and stream-based, time-dependent visualizations. This approach combines the advantages of both techniques and allows for novel visual layouts to emphasize hierarchical changes based on the application and user task. The visualization of static and dynamic hierarchies has been widely studied and inspired our approach.\\

\noindent
\textbf{General Hierarchies:} A multitude of techniques for the visualization of static hierarchies have been proposed in previous work. Treevis.net~\cite{schulz2011treevis} summarizes many of these approaches and provides search and filter functionality. The authors further categorize methods into explicit and implicit visualization techniques, as well as hybrid forms. Explicit methods are mostly based on a node-link diagram, where every data item is represented by a node (e.g., a circle) and relations among these nodes are presented by a link (e.g., a line connecting two circles). Implicit methods, on the other hand, are not required to draw links between nodes, but utilize positioning of individual nodes to represent hierarchical relationships. Treemaps~\cite{johnson1991tree} are one of the most influential examples of implicit hierarchy visualizations, representing nodes as rectangles and nesting child nodes within the space of their parent element. The space is vertically or horizontally split, creating rectangles proportionally sized to the numerical values of the child nodes.
Significant research efforts have been devoted towards optimizing many aspects of both implicit and explicit techniques, including considerations such as layout and aesthetics. One particular approach we build on are one-dimensional treemaps~\cite{johnson1993treemaps}, which nest nodes inside their parents, but always split the space along the same dimension. As demonstrated by ArcTrees~\cite{neumann2005arctrees}, this approach frees the second screen dimension to represent additional information, such as the temporal evolution of the hierarchy.\\

\noindent
\textbf{Juxtaposed Hierarchies:} For the visualization of hierarchies over time, one can in principle use any existing visualization method to display a static hierarchy in each timestep, either in a side-by-side (juxtaposed) manner or using animation. While TimeTree~\cite{card2006time} displays a node-link diagram and provides a slider to navigate through time, TreeJuxtaposer~\cite{munzner2003treejuxtaposer} integrates node-link diagrams in a juxtapositional manner. Isenberg and Carpendale~\cite{isenberg2007interactive} analyzed tree comparison tasks for juxtaposed tree layouts in an interactive multi-user setup. Several works have stabilized the positioning of individual data items in treemaps and optimized the layout strategy for easier comparison and tracking of items~\cite{tu2007visualizing, sud2010fast, tak2013enhanced, hahn2014visualization, van2017stable, sondag2018stable}. Vernier et al.~\cite{vernier2019quantitative} performed a quantitative comparison of 13 different treemap layouts and collected a benchmark dataset consisting of 2720 evolving hierarchies for that purpose.
While juxtaposed trees provide a good understanding of hierarchical structures at a given point in time, they lack a clear representation of time-related changes. They require the user to keep a mental map of all nodes and track their position and information encoding (e.g., color, size). Therefore, such techniques fail in conveying the evolution over time for larger trees and longer time series.\\

\noindent
\textbf{Static Hierarchies:} Several approaches deal with data where values associated with nodes or links change dynamically, but the hierarchy stays constant over the whole time span. SemaTime~\cite{stab2010sematime} and Timeline Trees\cite{burch2008timeline} display time-dependent information for each leaf node of a static hierarchy. Burch and Weiskopf~\cite{burch2011visualizing} visualize dynamic values along the links of connected nodes. Based on the ThemeRiver metaphor~\cite{havre2002themeriver}, and extensions like Byron and Wattenberg's geometry and aesthetic optimizations~\cite{byron2008stacked}, similar approaches have been developed to visualize data values over time together with their hierarchical structure. BookVoyager~\cite{wattenberg2006designing} displays the hierarchy explicitly, as an indented tree, separated from the stream-based visualization. TouchWave~\cite{baur2012touchwave} implicitly integrates the interaction with the hierarchy into a streamgraph. Both approaches can be utilized to hide individual streams for better readability and scalability. HierarchicalTopics~\cite{dou2013hierarchicaltopics} compute a meaningful hierarchy for non-hierarchical data to utilize the same interaction techniques. MultiStream~\cite{cuenca2018multistream} integrates multiple interaction methods to improve the focus and context awareness, enable linking and brushing, and tackle scalability issues with respect to time and hierarchical complexity. However, these types of methods are limited to the visualization of a static hierarchy. Our goal is to represent hierarchical changes over time in addition to the evolution of data values.\\

\noindent
\textbf{Dynamic Hierarchies:} In the tree-ring metaphor by Ther\'{o}n~\cite{theron2006hierarchical}, nodes of a tree are placed on concentric circles based on their date of addition to the tree, where larger rings correspond to later points in time.
Since streams provide an intuitive visual representation of changes over time, several methods adopt this approach to visualize dynamically changing hierarchies. Outflow~\cite{wongsuphasawat2012exploring} draws streams between nodes of individual timesteps and applies a hierarchical clustering to address scalability issues. Burch et al.~\cite{burch2014visualizing} draw an indented tree at each timestep and connect related nodes via links. Based on narrative charts~\cite{munroe2009movie} where elements split and merge over time, Textflow~\cite{cui2011textflow} visualizes hierarchical relationships as contours in the background of storylines. Tanahashi and Ma Cui~\cite{tanahashi2012design} optimize such layouts to create visually pleasing results and Liu et al~\cite{liu2013storyflow} further emphasize the representation of hierarchical structures. Cui et al.~\cite{cui2014hierarchical} introduce specific visual encodings to improve the understanding of hierarchical changes in stream-based visualizations, as well as tree cuts, which define the visibility of nodes at each time step.

\begin{figure}
  \begin{subfigure}{0.22\columnwidth}
    \centering\scalebox{1}[1.5]{\includegraphics[width=\columnwidth]{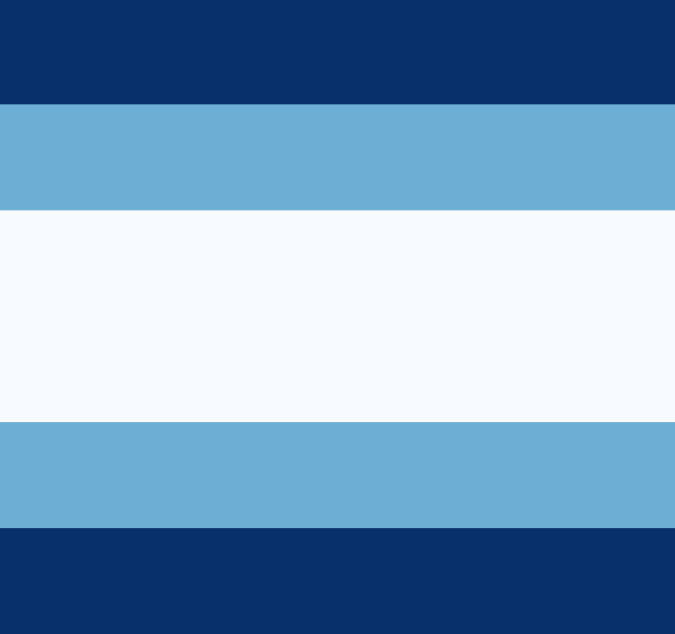}}
    \caption{}
  \end{subfigure}
  \hspace{3mm}
  \begin{subfigure}{0.44\columnwidth}
    \centering\scalebox{1}[1.5]{\includegraphics[width=\columnwidth]{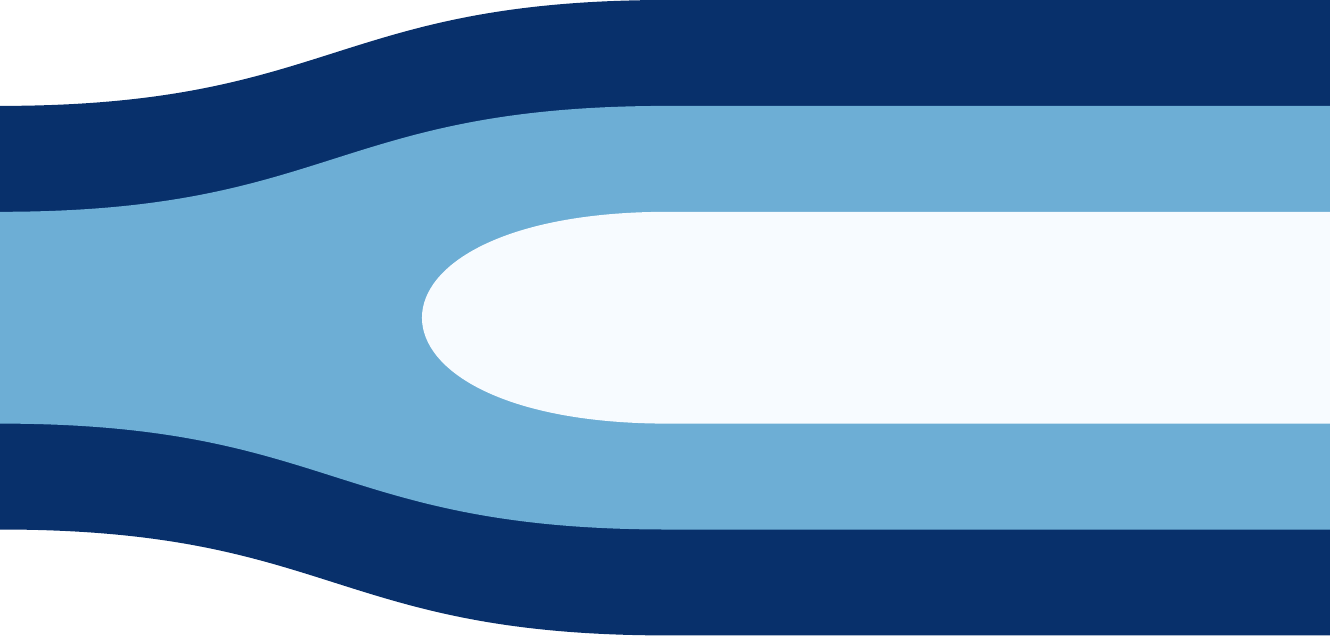}}
    \caption{}
  \end{subfigure}
  \hspace{3mm}
  \begin{subfigure}{0.22\columnwidth}
    \centering\scalebox{1}[1.5]{\includegraphics[width=\columnwidth]{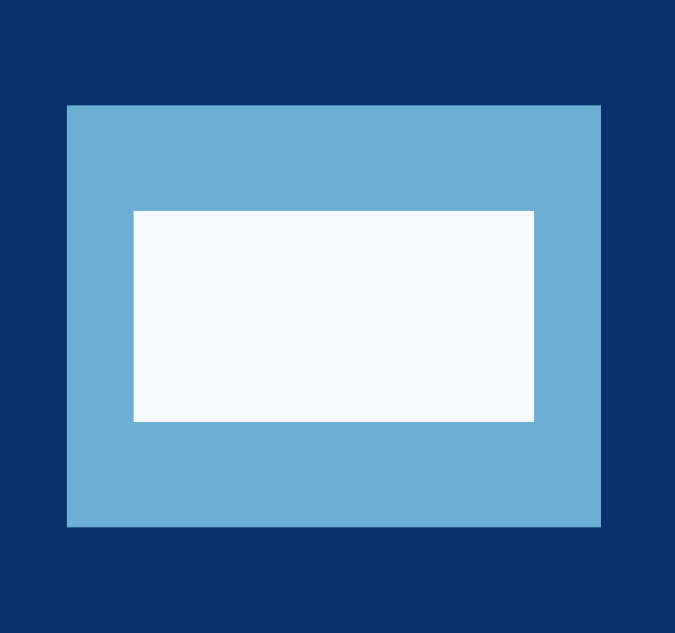}}
    \caption{}
  \end{subfigure}
  \caption{Perception of hierarchies. (a) If the hierarchy is only encoded by the nesting of streams, it is not clear if the visualization shows three streams (white contained by light blue, contained by dark blue) or five individual streams. (b) When a hierarchical change is present, the two-dimensional containment provides an intuitive understanding of the nested structures. (c) In our approach, we split the streams and introduce margins to the visualization, to clearly represent hierarchical structures at any given point in time.}
  \label{fig:explainProblem}
\end{figure}

Nested Tracking Graphs~\cite{lukasczyk2017nested} introduced a stream layout similar to ThemeRiver~\cite{havre2002themeriver} for hierarchically structured data by nesting streams inside each other. This representation allows for additions, deletions, merges, and splits in the hierarchy, but can only be applied to data where every node has a significantly larger value than the sum of values of its children. This requirement ensures that the hierarchical nesting is visible. Temporal treemaps~\cite{kopp2019temporal} extended this approach by a hierarchy-aware ordering for the reduction of edge crossings and visual cushions to support data where parents inherit their values from their children. The presented method is limited in the number of possible hierarchical changes, because only moves, splits, and merges of streams along siblings are taken into account. Movements across hierarchy levels, where a node changes its parent, are not visually represented. Chronicler~\cite{wittenhagen2016chronicler} visualizes the evolution of source code and
allows for the movement of nodes across the hierarchy. While all these stream-based techniques manage to visualize changes over time, they suffer from limited clarity in conveying the hierarchical structure at a given timestep. This issue becomes more apparent as the number of nodes or the number of timesteps increases. As demonstrated in \autoref{fig:explainProblem}, the main problem is that the hierarchical nesting of streams only becomes visible at times of hierarchical change. As long as no hierarchical change occurs, the number of individual streams and their respective nesting is only conveyed by color, but the interpretation of shapes can be ambiguous. In order to get a complete picture of the hierarchy at a certain timestep, at least one hierarchical change for every single stream needs to be visible, which negatively affects the readability when a large number of timesteps or nesting levels are present. We therefore propose a new visual metaphor which can not only reproduce these existing visualization techniques but can further emphasize the hierarchical structure at every single point in time without entirely relying on color coding to represent the depth of a node. 

%% file: s3_overview.tex
\begingroup
\setlength{\intextsep}{3pt}
\setlength{\columnsep}{6pt}

\section{Overview}
\label{sec:methods}

\subsection{Data}

We work with hierarchically structured data in which both the values and the underlying hierarchy may vary over time. For consistency with previous work, we refer to elements within the hierarchy (\textit{tree}) as \textit{nodes}. Each node has a value, exists for a certain period of time, and maintains a parent-child relationship (\textit{link}) to other nodes. Every node can have an arbitrary number of children, but at most one parent. Nodes which have the same parent are called \textit{siblings}. Nodes without children are referred to as \textit{leaves}. A node without a parent is called \textit{root} and defines the highest hierarchy level. In case several root nodes exist, we create an artificial root as the parent of all root nodes. The \textit{depth} of a node is the length of the shortest path between this node and the root. A \textit{hierarchy level} describes a set of all nodes with the same depth.

In order to visualize the mapping between nodes from one tree to the next, tree changes need to be defined. Every node of a tree therefore requires an identifier. This ID can be unique throughout all timesteps to clearly identify the existence of a node at each point in time. Alternatively, the ID can be unique to the tree of the current timestep only, in which case every node needs to store the ID of its predecessor and/or successor in addition to its own ID. Nodes that are added to the tree in a particular timestep can be identified by not having a predecessor. In the same manner, nodes which were deleted do not have a successor.

There are different ways of assigning scalar values to nodes. In some data, the value of a node is described as the sum of the values of its children, so that only leaf nodes contain a value. In the following, we will refer to the sum of values of a node's children as \textit{aggregate} and refer to a node which uses this aggregate as its value as \textit{aggregated node}. An example would be population data, where each person has a value of 1 and the population within a state is calculated by the sum of all people living in the state. The population of a country is then inherently defined as the sum of all state populations and so on, until the world population builds the root node, containing the number of living people as an aggregate. In other cases, the value of a node can exceed its aggregate and requires the definition of its own value. Instead of population, we could consider the area of buildings in a city as values. The area of a city district, which would represent the parent node, does not only contain buildings, but additional empty space and requires its own area value. In the same manner, states cover space which is not part of any city, and the world covers area which is not part of individual continents.



\subsection{Visual Encoding}

Nodes and their relationships may undergo many changes over time which need to be visually represented. Nodes can change their value, be added, split, merged, and deleted. They can further build new hierarchical relationships by changing their order among siblings or changing their parent. In the following, we will describe all possible modifications in detail and introduce their visual representation in our method. All higher-level operations on trees, like node swaps or rotations, can be represented by combinations of these cases.

\subsubsection{Content Change}

\begin{wrapfigure}{r}{0.12\textwidth}
    \includegraphics[width=0.12\textwidth]{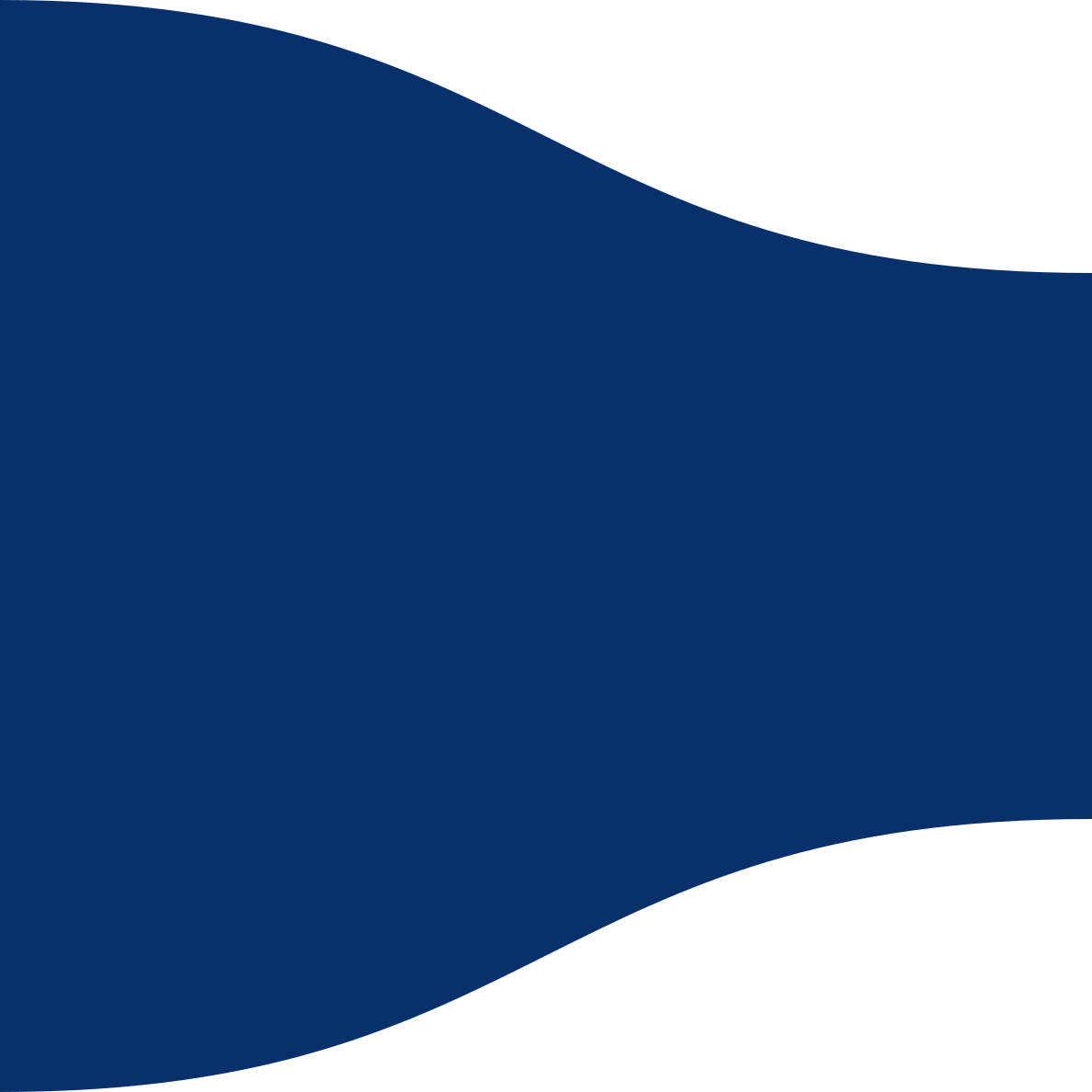}
\end{wrapfigure}
When the value of a node changes from one timestep to the next,
it is represented by a proportional change in height. The change
is visualized by an interpolation of the node's current representation and its position and value at the previous timestep. If values are only defined in leaf nodes of the tree, then the change in value propagates up and updates the values of ancestors in an iterative manner.

\begin{wrapfigure}{r}{0.12\textwidth}
    \includegraphics[width=0.12\textwidth]{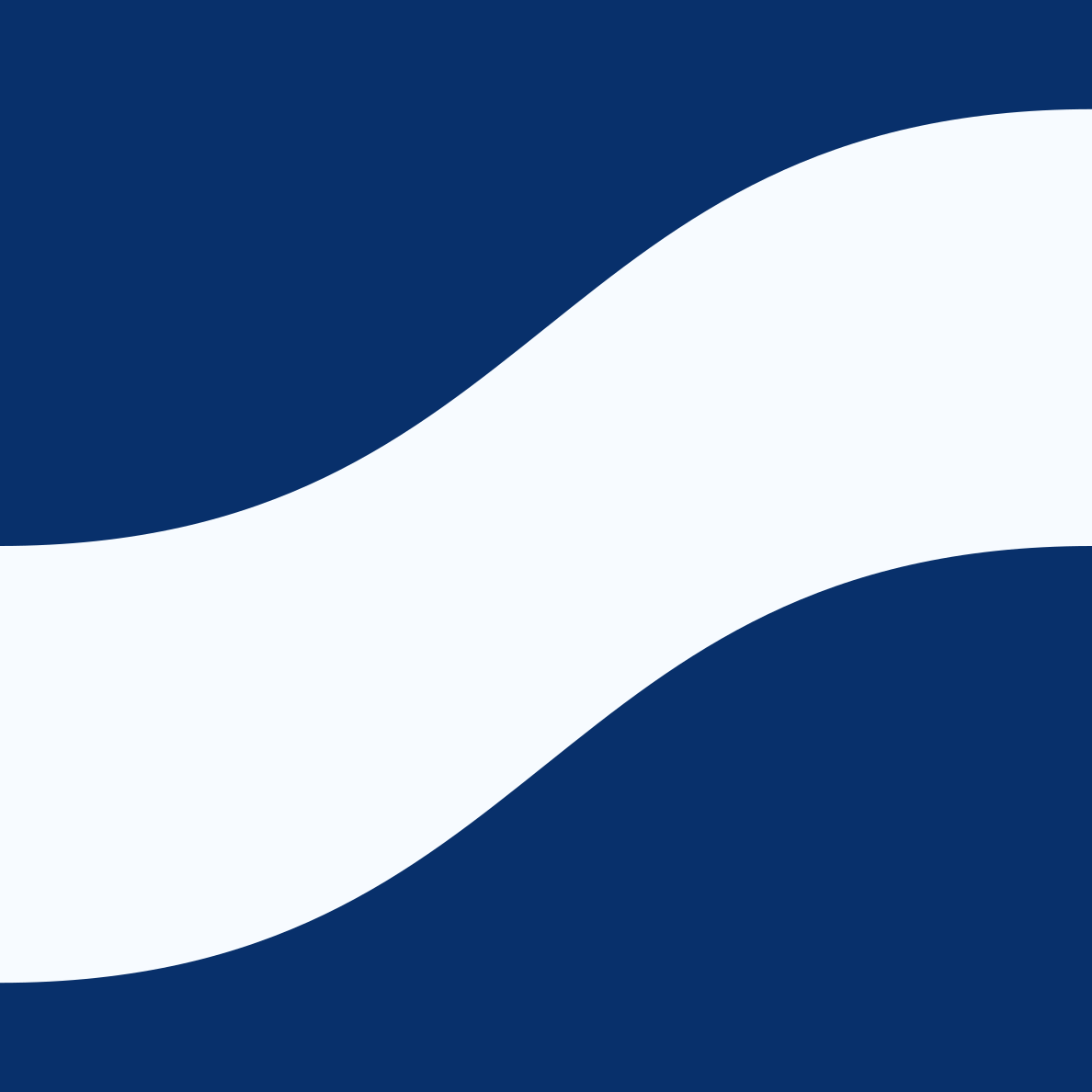}
\end{wrapfigure}
If the parent of a node has a higher value than its aggregate, then there is visual space available to move children without changing the sibling order. A node's position defines the distance of this node to the parent's starting point on the Y axis. The positional range by which a node can change while neither affecting hierarchical structure nor sibling order is limited by the position and size of the surrounding siblings.

Further changes to the content of a node can occur, e.g. in a document, where the text of a paragraph might change without changing its length. In this case the value and position of the node would stay unaffected and our visual representation would not change. While such changes can, for instance, be covered by changes in color encoding or tooltip information, they are specific to the underlying data and user task and will therefore not be discussed in this paper.

\subsubsection{Add \& Remove}

\begin{wrapfigure}{r}{0.12\textwidth}
    \includegraphics[width=0.12\textwidth]{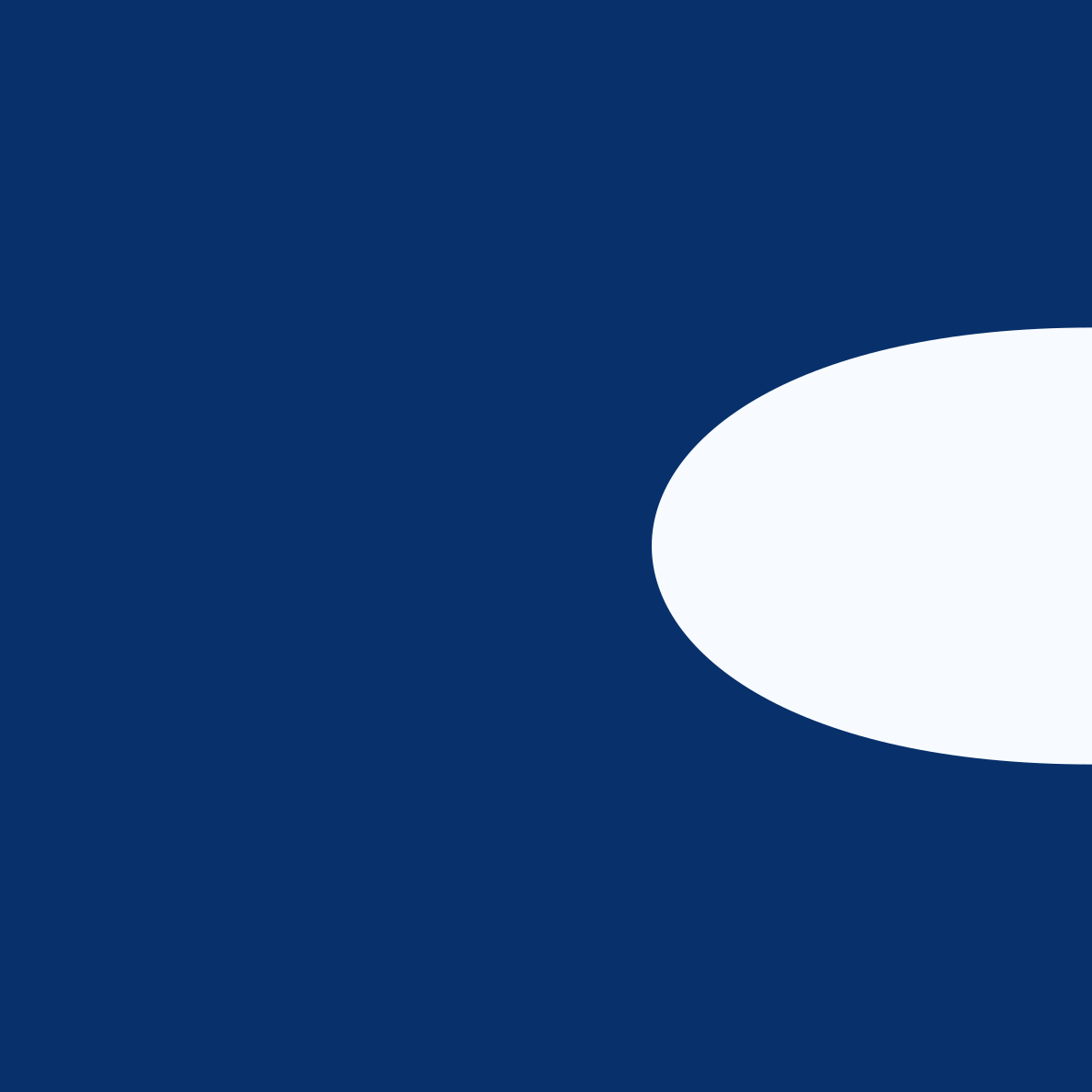}
\end{wrapfigure}
When a node is added to the data, we cannot interpolate between
the current timestep and the node's previous representation as it did not exist in the previous timestep. These cases could be visualized by interpolating between a zero value at the previous timestep and the value of the node at the current timestep. Since the node did not exist in the previous hierarchy, it is not clear where the zero value should be located. This is especially problematic when the hierarchical structure was very
different in the previous timestep. We therefore introduce a half ellipse in front of the stream to communicate the addition of nodes to the hierarchical structure, similar to the caps introduced by Chronicler~\cite{wittenhagen2016chronicler}. Analogously, the deletion of a node is represented by adding a half ellipse to the end of the stream.

\subsubsection{Split \& Merge}

\begin{wrapfigure}{r}{0.12\textwidth}
    \includegraphics[width=0.12\textwidth]{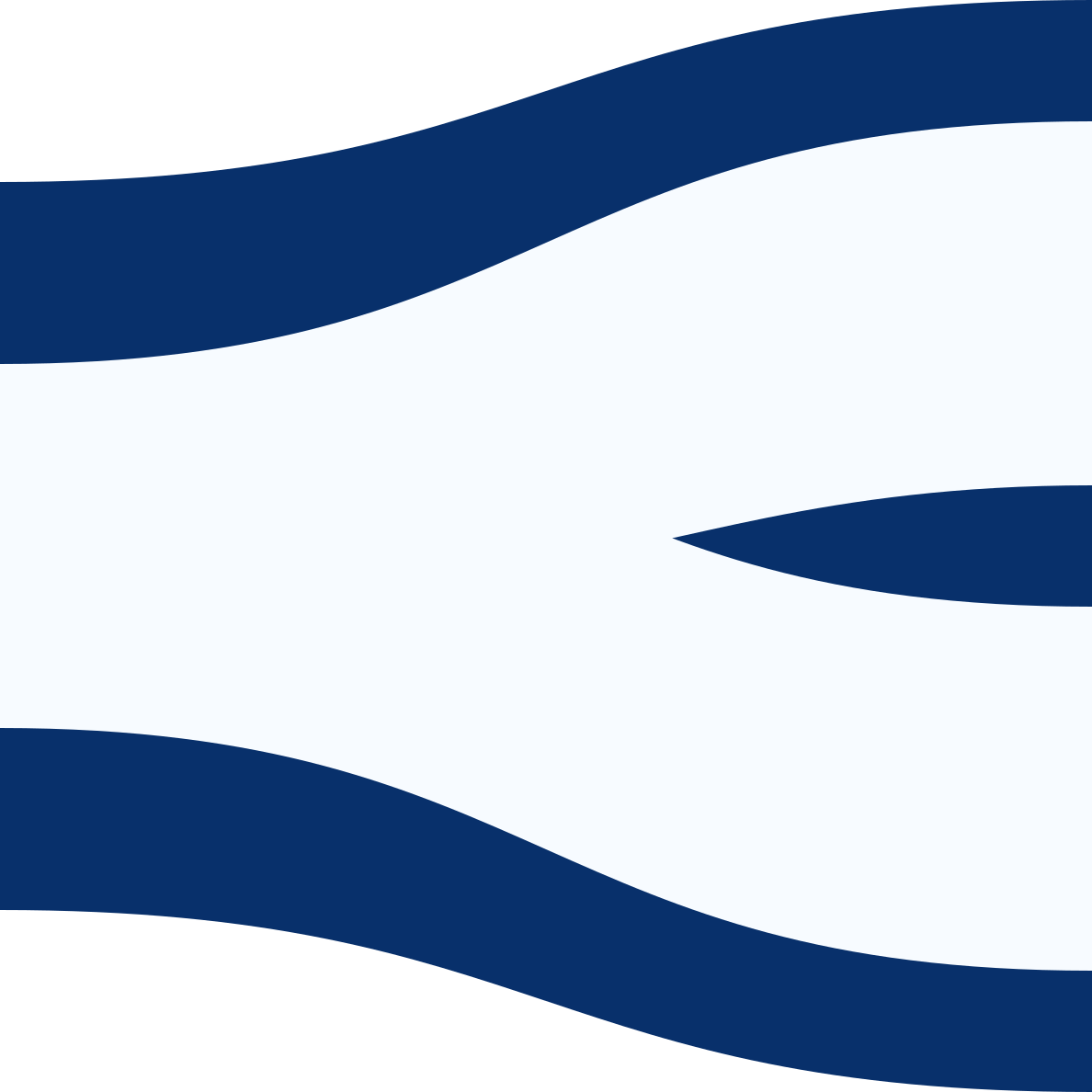}
\end{wrapfigure}
If a node has multiple predecessors, it means that multiple nodes merged into a single node from one timestep to another. If a node splits, it has multiple successors. The split in the data is visually represented by a split of the stream. The stream begins with the representation of the previous node and then splits into the representations of all involved nodes at the current timestep. In the same manner, merges of nodes are represented by a merge of individual streams into a single one. Depending on the locations of merges and splits, stream crossings might be inevitable \cite{kopp2019temporal}.

\subsubsection{Move}

We will now discuss move operations that always introduce stream crossings to the visualization. We differentiate between three cases: Movement within a node, across nodes, and along ancestors, which can be seen as a special case of the former.\\

\noindent
\begin{wrapfigure}{r}{0.12\textwidth}
    \includegraphics[width=0.12\textwidth]{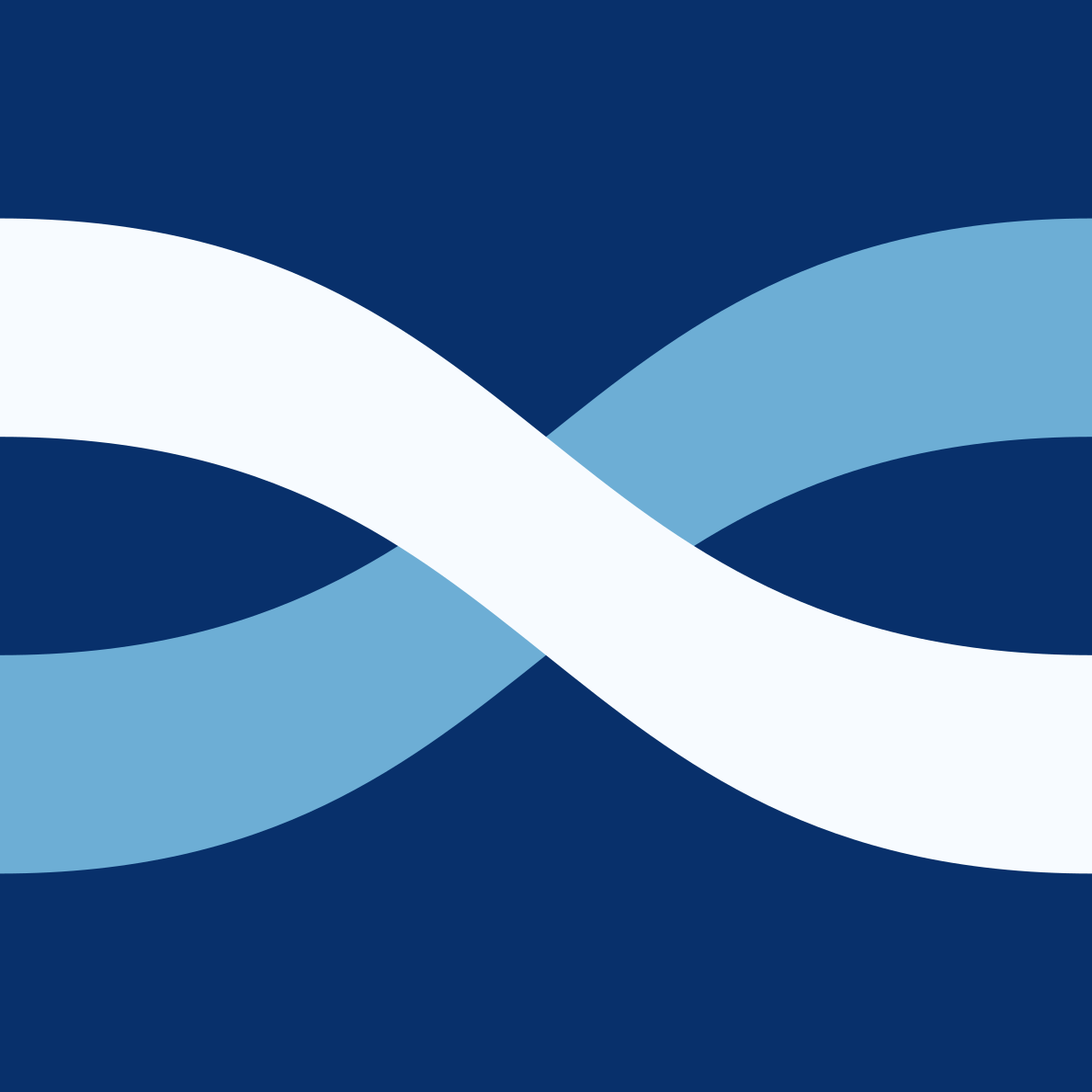}
\end{wrapfigure}
\textbf{Movement Within a Node:}
Movement within a node is defined by a reordering of a node's children. This type of movement can only occur when there is a predefined order to the tree nodes. As an example, a tree in one timestep might consist of a root node with two children A and B. In the next timestep, the positioning of the children has switched, so that they are stored in the order (B,A) within the root node. When interpolating between nodes at their respective positions, the child streams will cross each other. The number of edge crossings within the dataset largely depends on the ordering. If the order of children does not matter for the represented data, the number of edge crossings can be reduced by applying a suitable sorting algorithm.\\

\noindent
\begin{wrapfigure}{r}{0.12\textwidth}
    \includegraphics[width=0.12\textwidth]{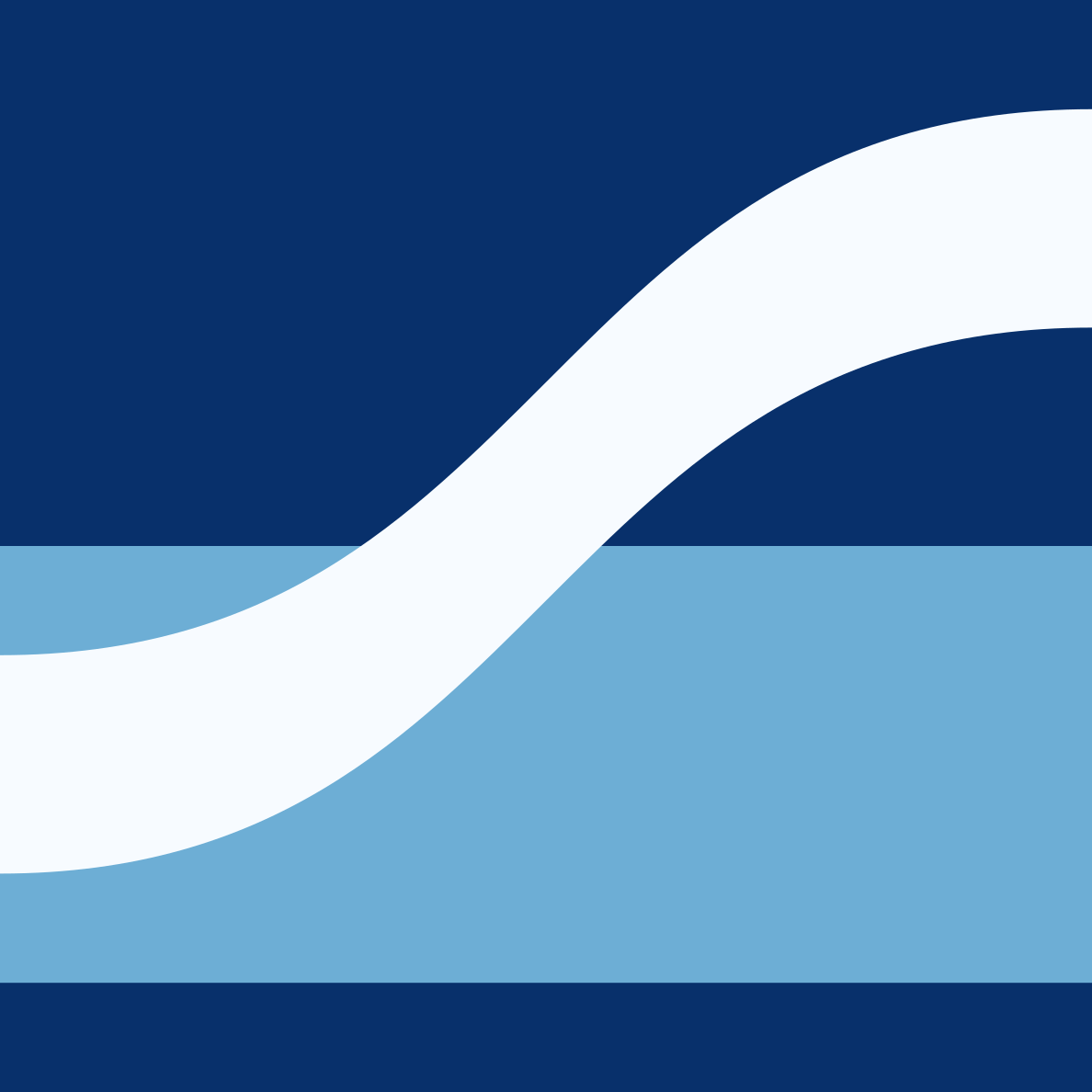}
\end{wrapfigure}
\textbf{Movement Across Nodes:} We define a node changing its parent from one timestep to the next as a hierarchical change across nodes. Since our visualization is built around nested streams, the stream representing such a structural change will at least have to cross the border of its parent element. Depending on the depth difference between the nodes and based on the current order of the tree, the stream might cross several other streams.\\

\noindent
\begin{wrapfigure}{r}{0.12\textwidth}
    \includegraphics[width=0.12\textwidth]{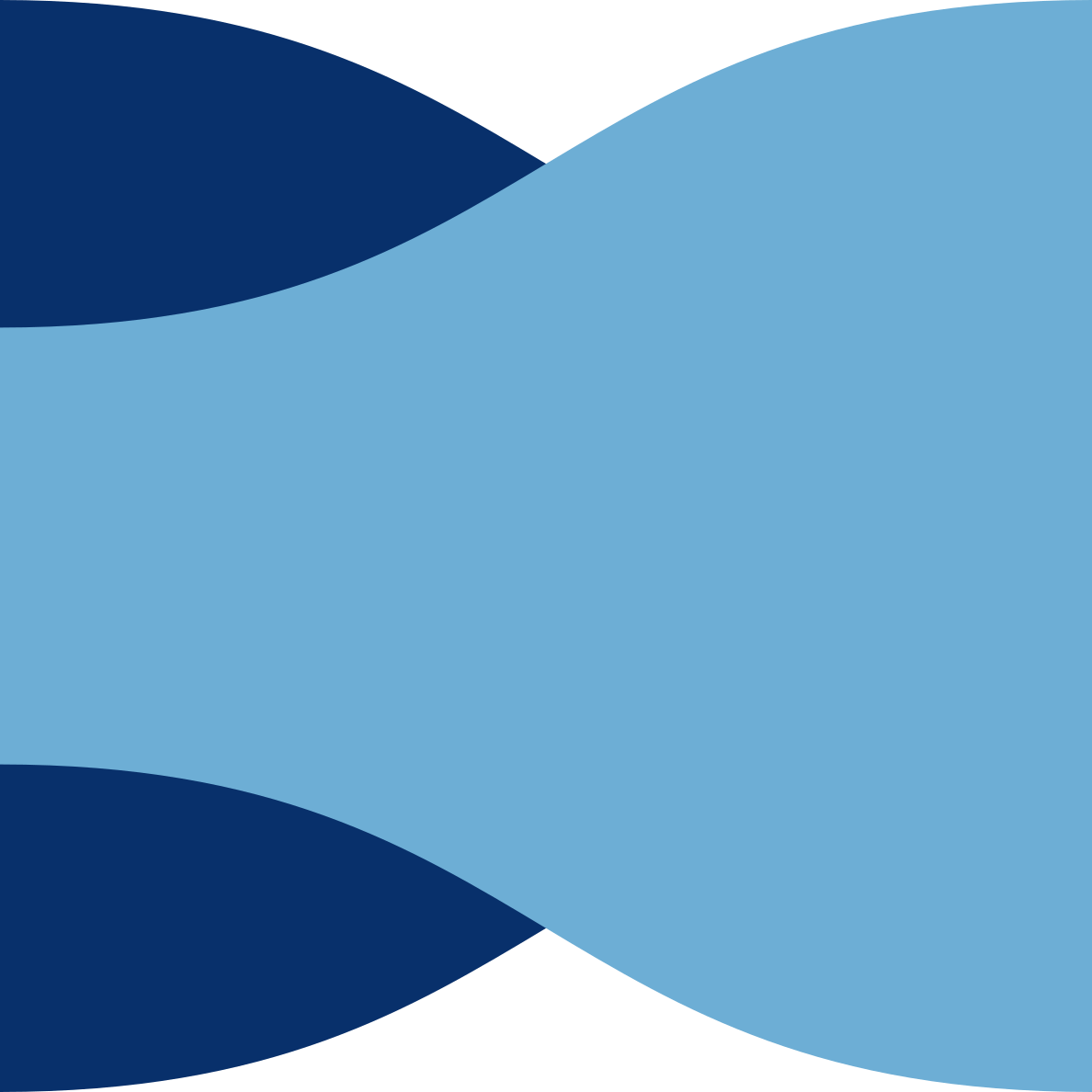}
\end{wrapfigure}
\textbf{Movement Along Ancestors:} A special case of nodes moving across hierarchy levels is nodes becoming the parent of one of their previous ancestors. Let us assume a tree that only consists of a root node with a single child. If these two nodes switch their position from one timestep to the other, then the child becomes the parent of its previous parent.
Such a scenario could for example occur when switching the inner and outer loop of an algorithm during development.
The two structural nodes would change their position and swap the hierarchical level they live in. The problem that occurs lies in the drawing of streams, where streams of higher depth are always drawn on top of the streams they are nested in. Since, in this case, the nesting changes, we would need to draw the child on top of its parent in one timestep, but switch this order for the next timestep. In order to solve this conflict, we are required to cut one of the two streams in half and draw the first half on top and the second half underneath the other stream or vice versa.

\begin{wrapfigure}{r}{0.12\textwidth}
    \includegraphics[width=0.12\textwidth]{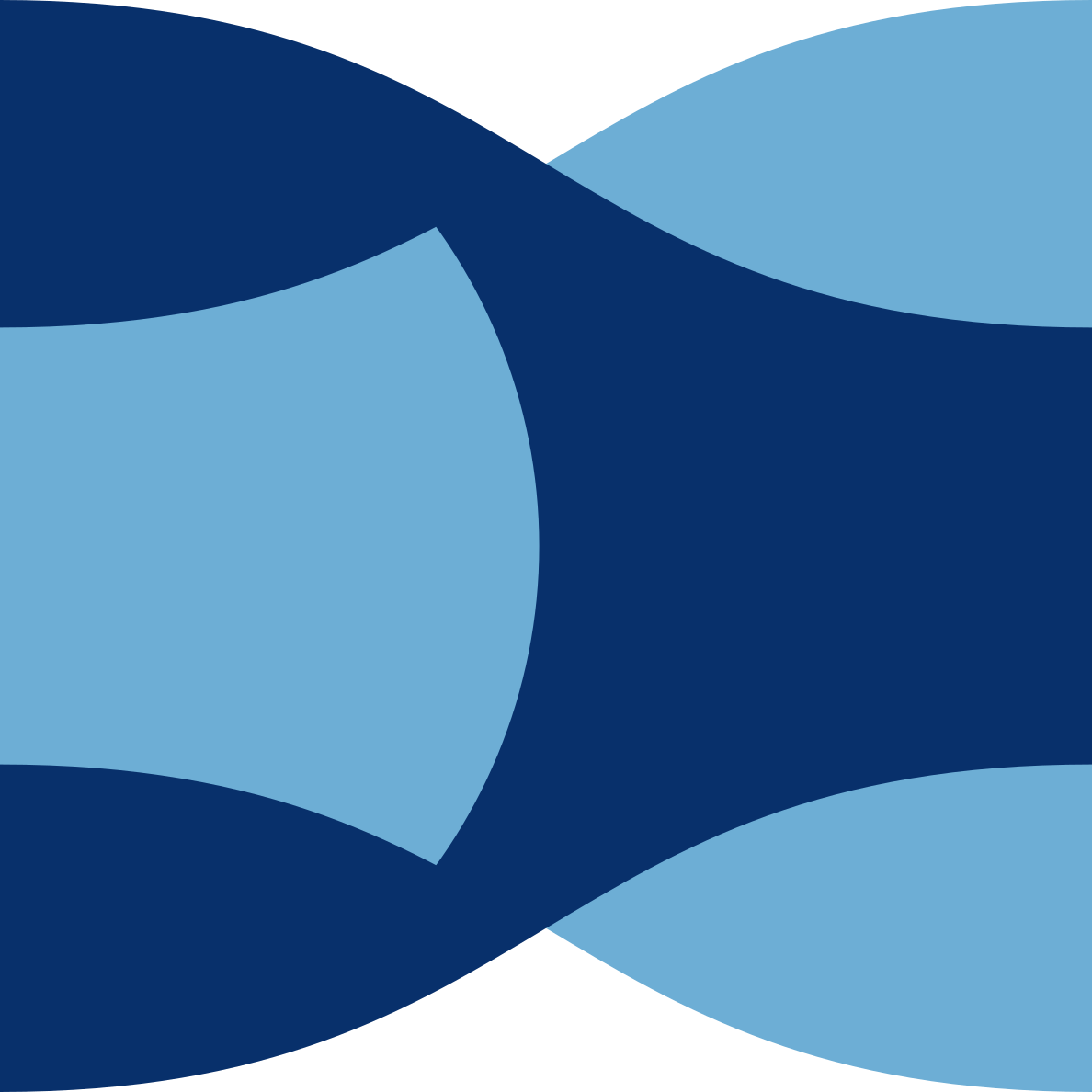}
\end{wrapfigure}
To come up with a visually more pleasing solution, we imagine the streams as two sheets lying on top of each other, cut a hole into one of the streams, and thread the other through the hole. The technique scales to an arbitrary number of such movements by adding more holes and threads to the streams. This representation is not only required when a child node switches positions with its parent, but whenever a node becomes an ancestor of any of its previous ancestors. The detection of such cases requires a complete tree traversal.

\endgroup 

%% file: s4_visualDesign.tex
\section{SplitStream Generation}
\label{sec:visualDesign}

In the following section we describe our metaphor for visualizing evolving hierarchies. We will demonstrate how the general definition of our method allows for the smooth transition between one-dimensional treemaps and nested stream visualizations, as well as the generation of our own visual approach. We further provide a detailed explanation of parameters to adjust the visualization based on the data and task at hand. 

\subsection{Hierarchy-Change Ratio}

\begin{figure}
  \begin{subfigure}{\columnwidth}
    \centering\scalebox{1.85}[1]{\includegraphics[height=2.9cm]{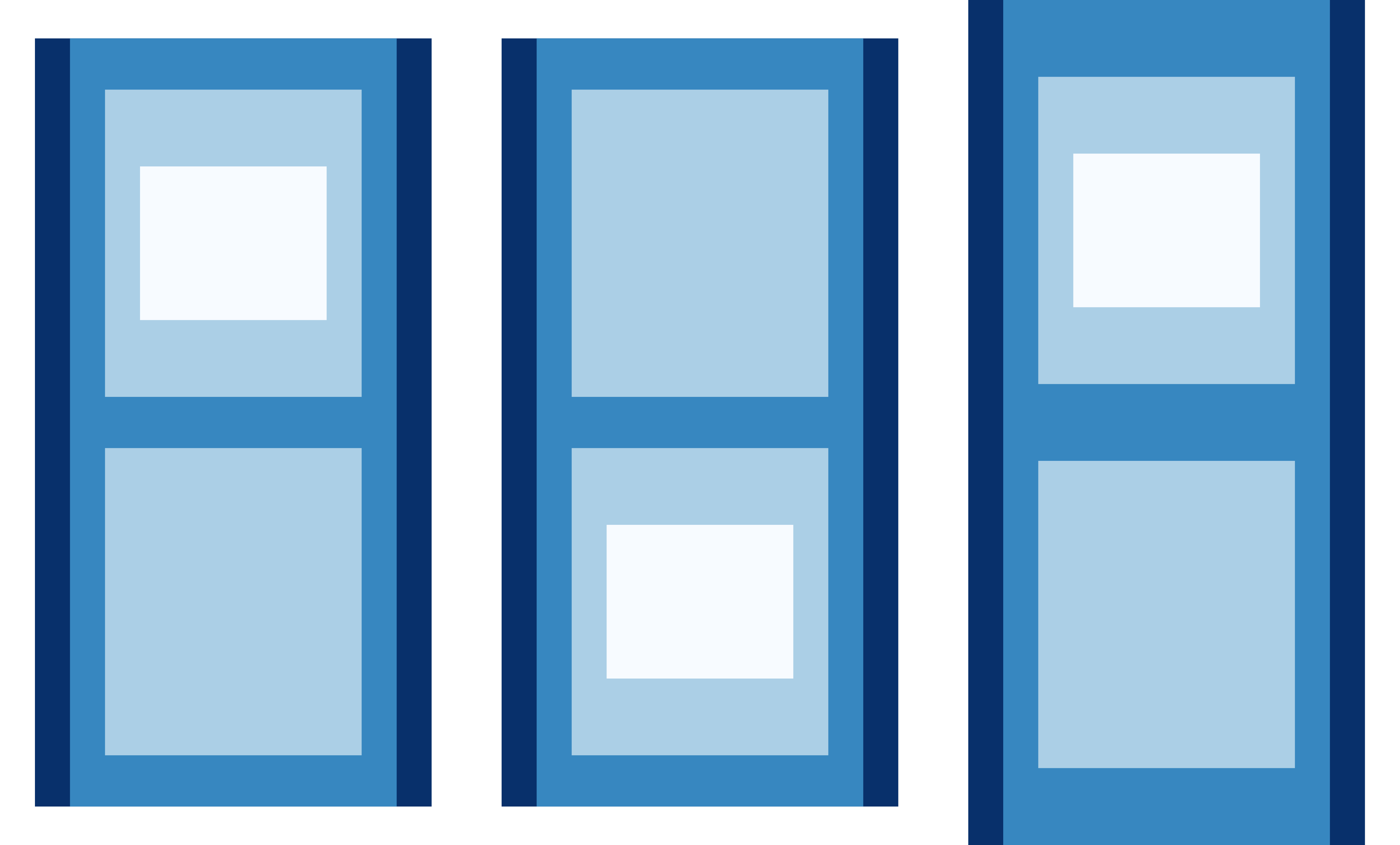}}
    \caption{One-dimensional treemap over time. The white rectangle changes its parent element from step~1 to step 2. Between step~2 and step~3, the white node is deleted and a new white node is created. These hierarchical changes are not visually distinguishable.}
    \label{fig:treemap}
    \vspace*{3mm}
  \end{subfigure}
  \begin{subfigure}{\columnwidth}
    \centering\scalebox{1.85}[1]{\includegraphics[height=2.9cm]{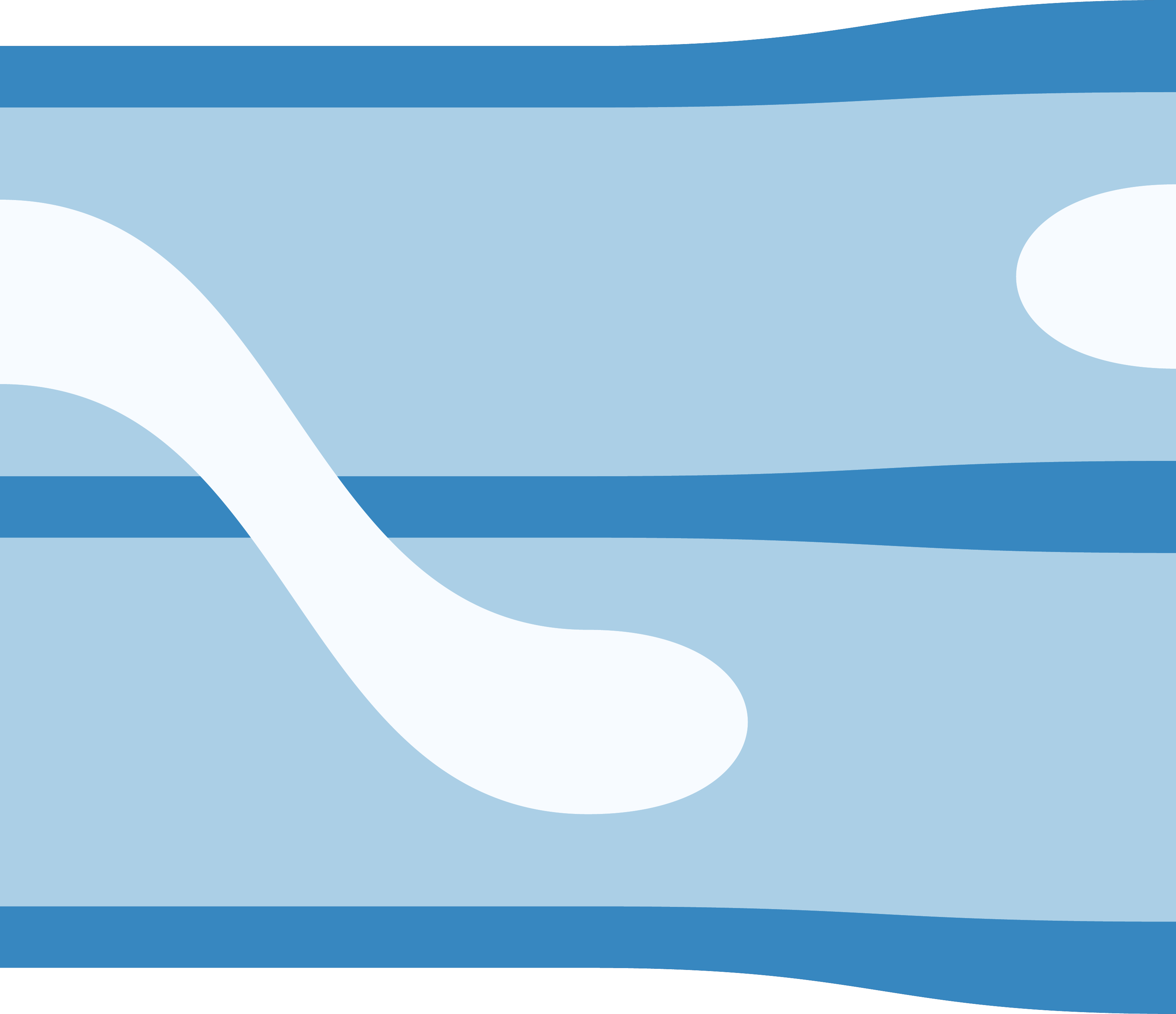}}
    \caption{Nested streamgraph. Hierarchical changes are clearly visualized, but the structural nesting is merely encoded in color. Aggregated nodes, such as the dark blue root node, are not visible.}
    \label{fig:stream}
    \vspace*{3mm}
  \end{subfigure}
  \begin{subfigure}{\columnwidth}
    \centering\scalebox{1.85}[1]{\includegraphics[height=2.9cm]{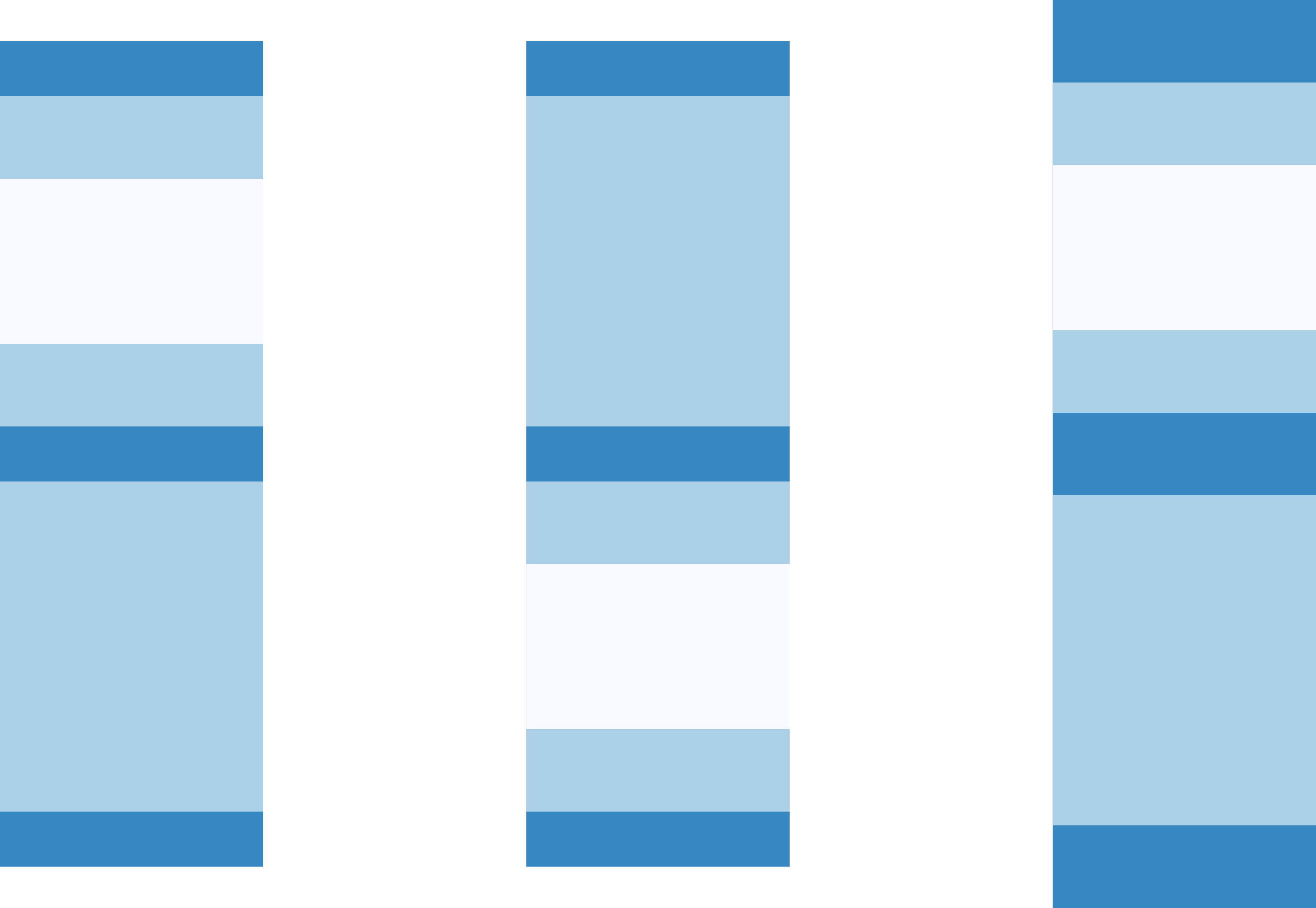}}
    \caption{One-dimensional treemap over time with a fixed width for each node.}
    \label{fig:secstream1}
    \vspace*{3mm}
  \end{subfigure}
  \begin{subfigure}{\columnwidth}
    \centering\scalebox{1.85}[1]{\includegraphics[height=2.9cm]{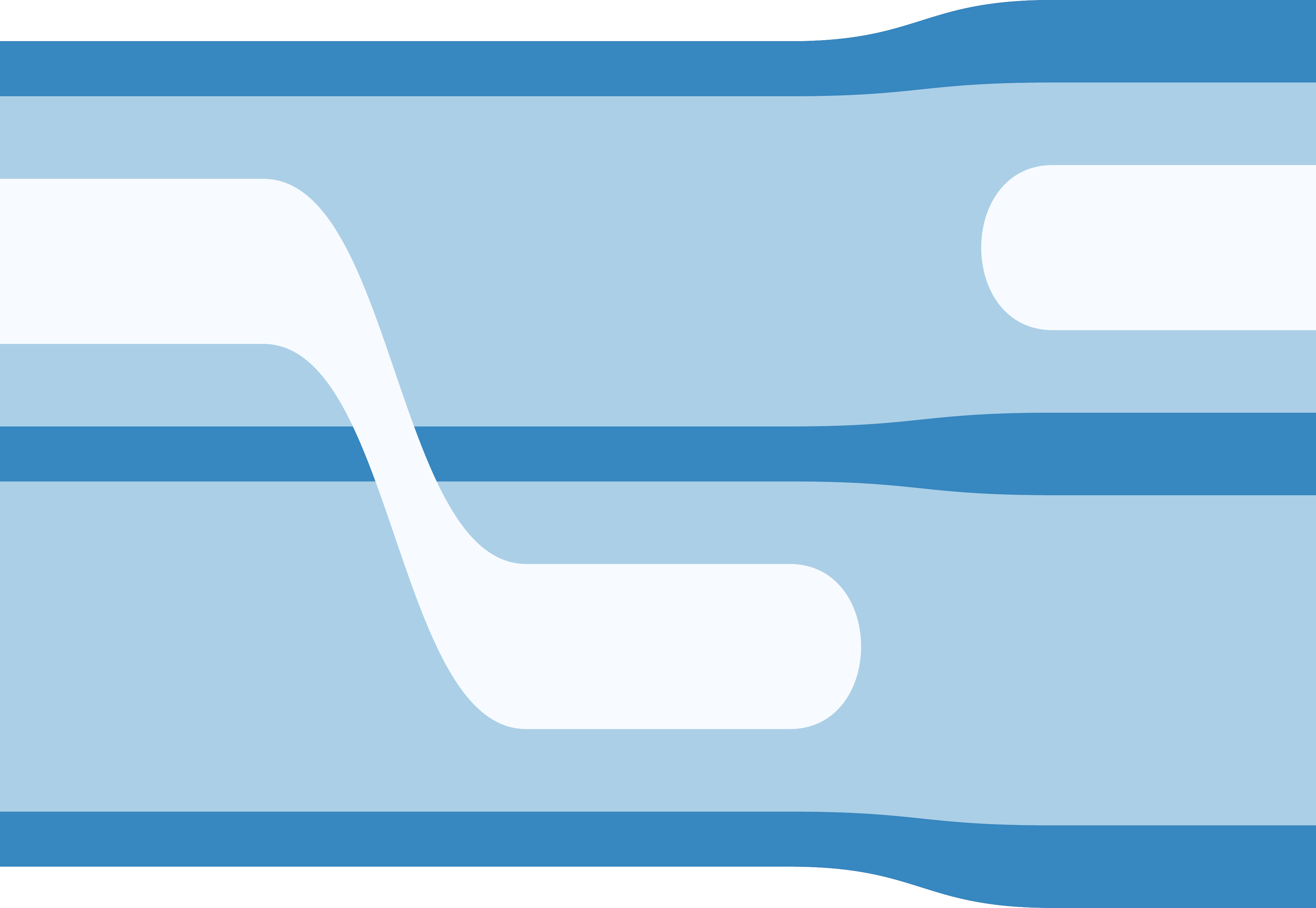}}
    \caption{Connecting the nodes of individual treemaps by streams, to represent hierarchical changes.}
    \label{fig:secstream2}
    \vspace*{3mm}
  \end{subfigure}
  \begin{subfigure}{\columnwidth}
    \centering\scalebox{1.85}[1]{\includegraphics[height=2.9cm]{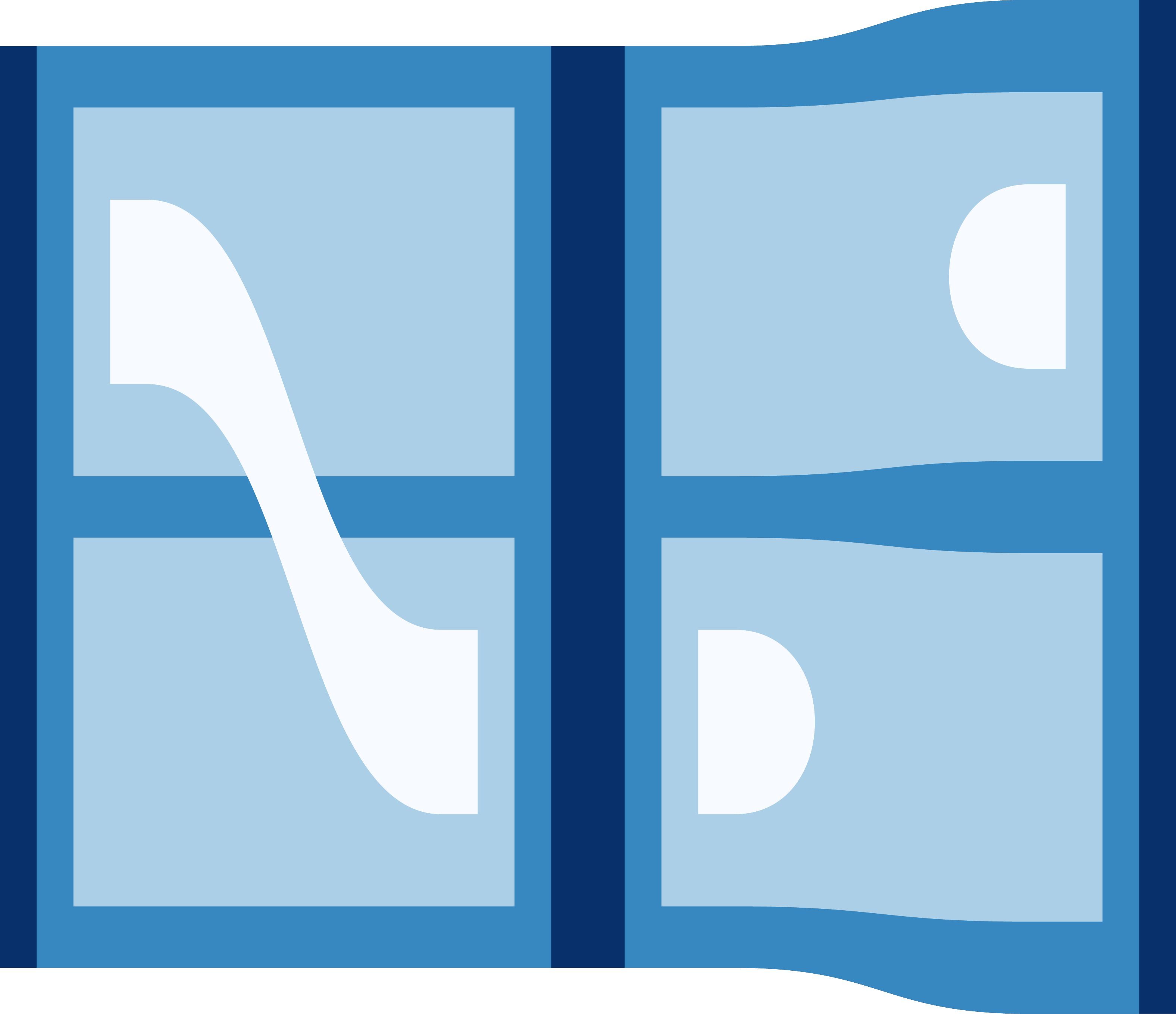}}
    \caption{SplitStreams. Each treemap is split in the center and all streams are inset based on their hierarchy level. The hierarchical changes over time, the nested structures at each point in time, and aggregated nodes are now clearly visible.}
    \label{fig:secstream3}
  \end{subfigure}
  \caption{Treemaps~(a) visualize hierarchies at specific timesteps and nested streamgraphs~(b) display structural changes over time. SecStreams~(e) combine these approaches to accomplish both tasks in one static visualization. All depicted visualizations share the same dataset and a common time axis.}
  \label{fig:generation}
\end{figure}


When visualizing hierarchies over time, one can display a static tree representation for each timestep in a juxtaposed manner. One example would be to compute a treemap for each timestep and show all these treemaps next to each other sorted by time. 
In the following, we will only focus on one-dimensional treemaps, which draw each node as a rectangle, stack all children of a node on top of each other and nest them inside their parent element (\autoref{fig:treemap}). The height of each rectangle corresponds to the value of the node. The hierarchy is represented by assigning a smaller width to deeper nodes. This representation provides a clear depiction of the hierarchy even for aggregated nodes. The main issue when showing static hierarchies per timestep is that changes from one step to another are not readily apparent. The user needs to track the elements' position and color, which becomes more difficult as the number of elements and changes increases. Furthermore, some hierarchical changes cannot be distinguished at all. When a node N changes its parent, its rectangle will be nested inside its parent A in one timestep and then move to be nested inside its new parent B. The very same visualization would be created if N is deleted in A and a new child M is added to B. We can therefore not distinguish between a move operation and the combination of a delete and an add operation (\autoref{fig:treemap}).

Nested streamgraphs are trying to tackle this problem by visualizing hierarchical changes directly. One can think of this as drawing one-dimensional treemaps at each point in time, setting their width to 0, so that every treemap only consists of a single vertical line, and connecting the representations of each node at every timestep by a curve (\autoref{fig:stream}). Because the changes themselves are displayed, we can now easily distinguish the movement of a node from a delete and add operation. Furthermore, the comparison of values is now guided by connected lines and not dependent on the comparison of heights alone. The main problem we face when using streams instead of treemaps is that the hierarchical nesting is not explicitly visualized. The margins that used to display the nesting in treemaps were removed in favor of drawing continuous streams. While color can be utilized to represent hierarchy levels, the number of distinguishable shades of a color limits the levels of hierarchy we can display. Additionally, when a node changes its hierarchy level, it is unclear what color this node should be assigned. On top of that, aggregated nodes are not visible anymore, as shown in \autoref{fig:stream}.

We introduce SplitStreams, a hybrid approach which is meant to combine the benefits of both treemaps and nested streams, while reducing the drawbacks of both approaches. The goal is to visualize hierarchical changes over time directly, while still conveying the nesting of nodes at all timesteps. In order to accomplish this task, we draw a treemap at every point in time and set the width of all nodes to a fixed value (\autoref{fig:secstream1}). We then connect nodes between treemaps via streams, which leaves us with a visualization similar to the nested streams, but with horizontal lines in the treemap areas (\autoref{fig:secstream2}). Finally, we split the graph at every timepoint, in the center of each treemap, and move every stream by a certain \textit{margin} away from the split position. The margin that is applied to each node is based on its hierarchical level, which reintroduces the representation of nested structures (\autoref{fig:secstream3}). What we end up with is a visualization of changes, where each block displays the hierarchical structure at two points in time, one in the beginning and one in the end. The part in the middle displays the change between both hierarchies.

In order to draw both, the treemaps at each timestep and the streams between them, we must divide the available space between two points in time and reserve a certain portion of it for each visualization method. If we reserve more space for the treemap, hierarchical changes will be more cramped and less visible. If we dedicate more space to be used by the stream representation, there will be less space for introducing margins and displaying nested structures at a given point in time. We call this trade-off the \textit{hierarchy-change ratio} (HCR). An HCR of 1 means that we are only showing hierarchy, a treemap at each point in time. An HCR of 0 will only represent change, leading to a nested streamgraph. This parameter allows for the smooth transition between one-dimensional treemaps and nested streamgraphs.

\subsection{Splits and X-Margins}

Treemaps utilize two spatial dimensions to visualize hierarchical nesting, so that rectangles of nodes are completely contained by the rectangle of their parent node. Nested streamgraphs replace one of these spatial dimensions to represent time, which helps to visualize changes, but reduces the user's capability to read hierarchical structure at a given timestep. Nodes of higher hierarchy levels are occluded by the continuous representation of their children. In order to visualize hierarchical structures, we introduce splits to cut the stream representation. We then define an x-margin, which opens up each split, by clipping the width of individual streams. When the width of a stream is reduced, its underlying streams become visible. By choosing a margin for each stream based on its hierarchy level (depth in a tree), we can display the complete hierarchical structure present at that point in time. One should keep in mind, that if the margin becomes too large and the space provided for each point in time is too small, nodes with high depth values will not be visible anymore. We therefore propose to provide screen space for each timestep based on the following formula:

\vspace{-4mm}
\begin{equation}
\label{equ:HCR}
HCR \cdot dist(t_i, t_{i+1}) > m_x\big(d_{max}(t_i)\big) + m_x\big(d_{max}(t_{i+1})\big)
\end{equation}
where $dist$ is the distance between two points in time on the x-axis, $d_{max}$ is the maximum depth of the hierarchy at time $t$ and $m_x$ is the margin defined for the depth given. We hereby assure that the distance between two timepoints is large enough to represent every single node given the desired margin. By adding the HCR to the equation, we further ensure that the margin will only be applied in the treemap areas, where streams are drawn as horizontal lines, so that streams will not change their starting and end position when being clipped. In order to keep the time axis linear, the distance between all timesteps should be fixed and larger than the maximum distance required to display the margins without reducing any node's width to zero.

\begin{figure}
    \centering\scalebox{1}[1]{\includegraphics[height=3.5cm]{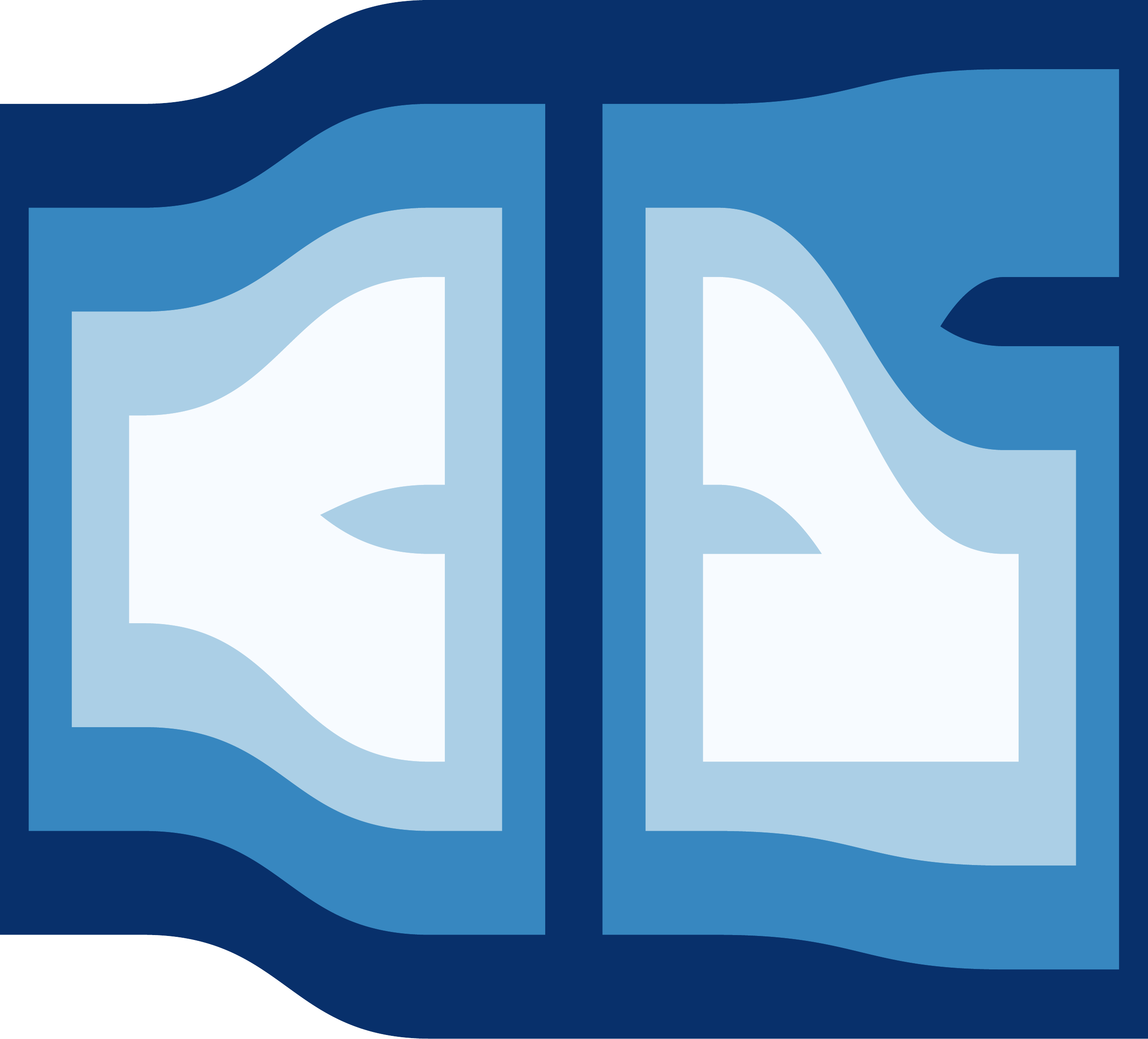}}
\hspace{0.3cm}
    \centering\scalebox{1}[1]{\includegraphics[height=3.5cm]{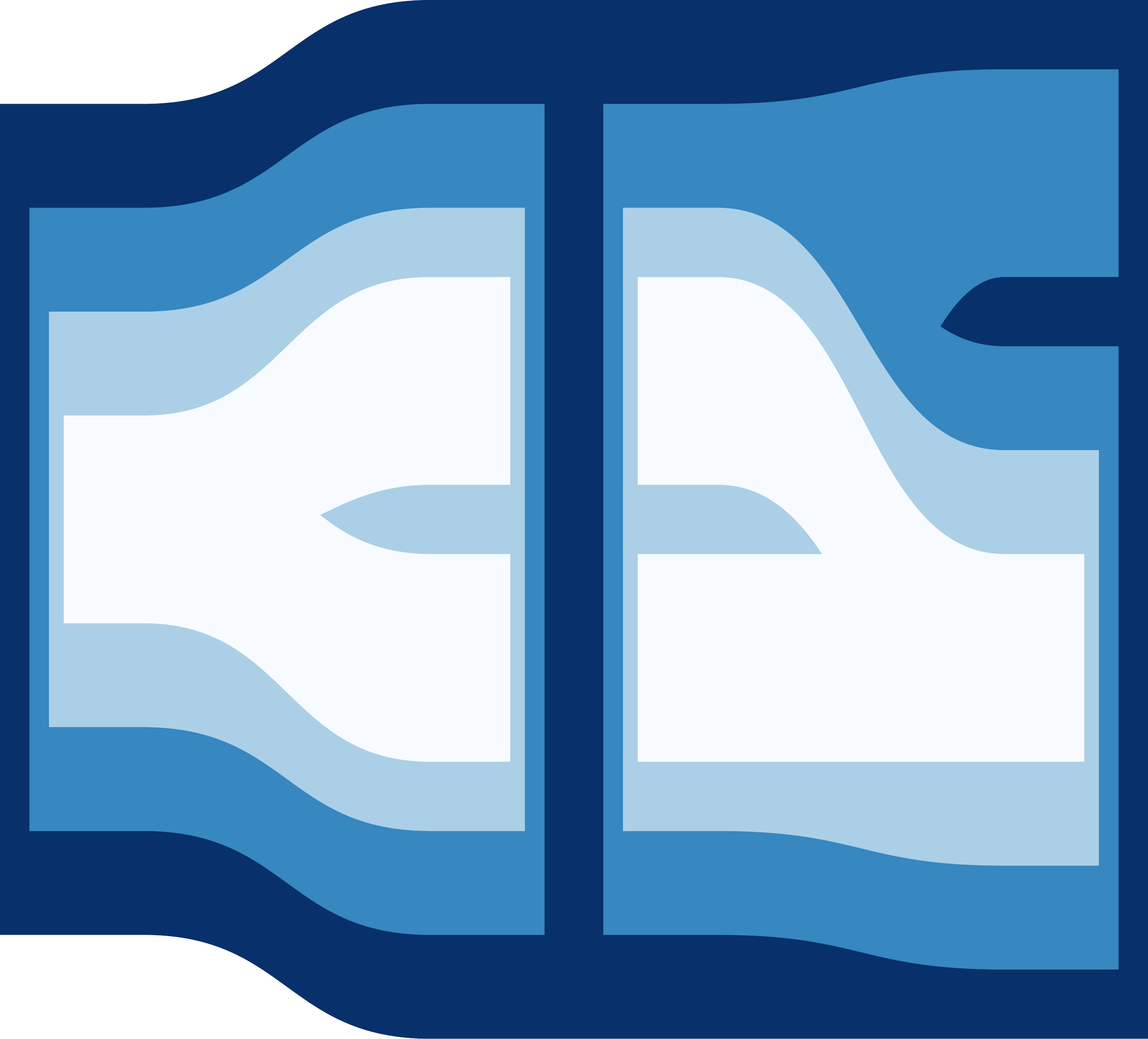}}
    \caption{Left: Hierarchical Margin. Streams receive a higher margin, the deeper their hierarchy level is. Right: Reversed Hierarchical Margin. Streams closer to the root receive larger margins. Although both cases feature the same margin on the root node, the latter case provides a clearer representation of changes over time, because leaf nodes are not pushed so far away from each other.}
    \label{fig:margin}
\end{figure}

For most of our examples, we utilize a fixed x-margin, which linearly decreases the width of nodes along the depth of the hierarchy by a fixed amount (\autoref{eq:fixed}). Depending on the task the visualization is supposed to solve, a non-linear scaling of margins might be better suited. The margins can for instance be steered to focus on a certain level of hierarchy by applying a bigger margin to nodes of a specific depth. By reducing a stream's width the deeper it is in the hierarchy (\autoref{eq:hierarchical}), we can emphasize hierarchical structures for deeper nodes (\autoref{fig:margin} left). When choosing a higher width reduction for nodes closer to the root (\autoref{eq:hierarchicalReversed}), leaf nodes are barely separated from their parents, keeping the appearance of streams as continuous as possible (\autoref{fig:margin} right). In order to avoid introducing overlaps between streams, we define margins in a recursive manner, so that every node's width is at least as much reduced as the width of its parent element. The margins can be represented as functions:

\vspace{-2mm}
\begin{equation}\label{eq:fixed}
m_{fixed} = m_x\big(p(n)\big) + value
\end{equation}

\vspace{-4mm}
\begin{equation}\label{eq:hierarchical}
m_{hier} = m_x\big(p(n)\big) + d(n) \cdot value
\end{equation}

\vspace{-4mm}
\begin{equation}\label{eq:hierarchicalReversed}
m_{hier^{-1}} = m_x\big(p(n)\big) + \frac{value}{d(n)}
\end{equation}
where $m_x$ is the margin, $n$ is a node in the tree at timepoint $t$, $p$ is its parent, $d$ is the node's depth, and $value$ is a fixed number which steers the size of the margin. For the root, which does not have a parent node, we set the margin to 0. It should be noted that a single stream can have different margins at different points in time, for instance if it moves into a new parent and thus changes its hierarchy level.

While both treemaps and SplitStreams utilize splits and margins to visualize nested structures, there are two notable differences in their generation. First, to keep the appearance of a continuous stream in SplitStreams, we do not introduce a margin for the root node. Second, as can be seen in \autoref{fig:generation}, splits for treemaps and SplitStreams are located at different positions on the time axis. For treemaps, we split streams between timesteps to separate them from each other. In our method, we split streams at timesteps to separate the changes. Although splits could be located at any possible position on the time axis, we did not encounter any meaningful examples for other split positions.

\subsection{Y-Padding and Y-Margin}

\begin{figure}
    \centering\scalebox{1}[1]{
        \begin{overpic}[height=3cm]{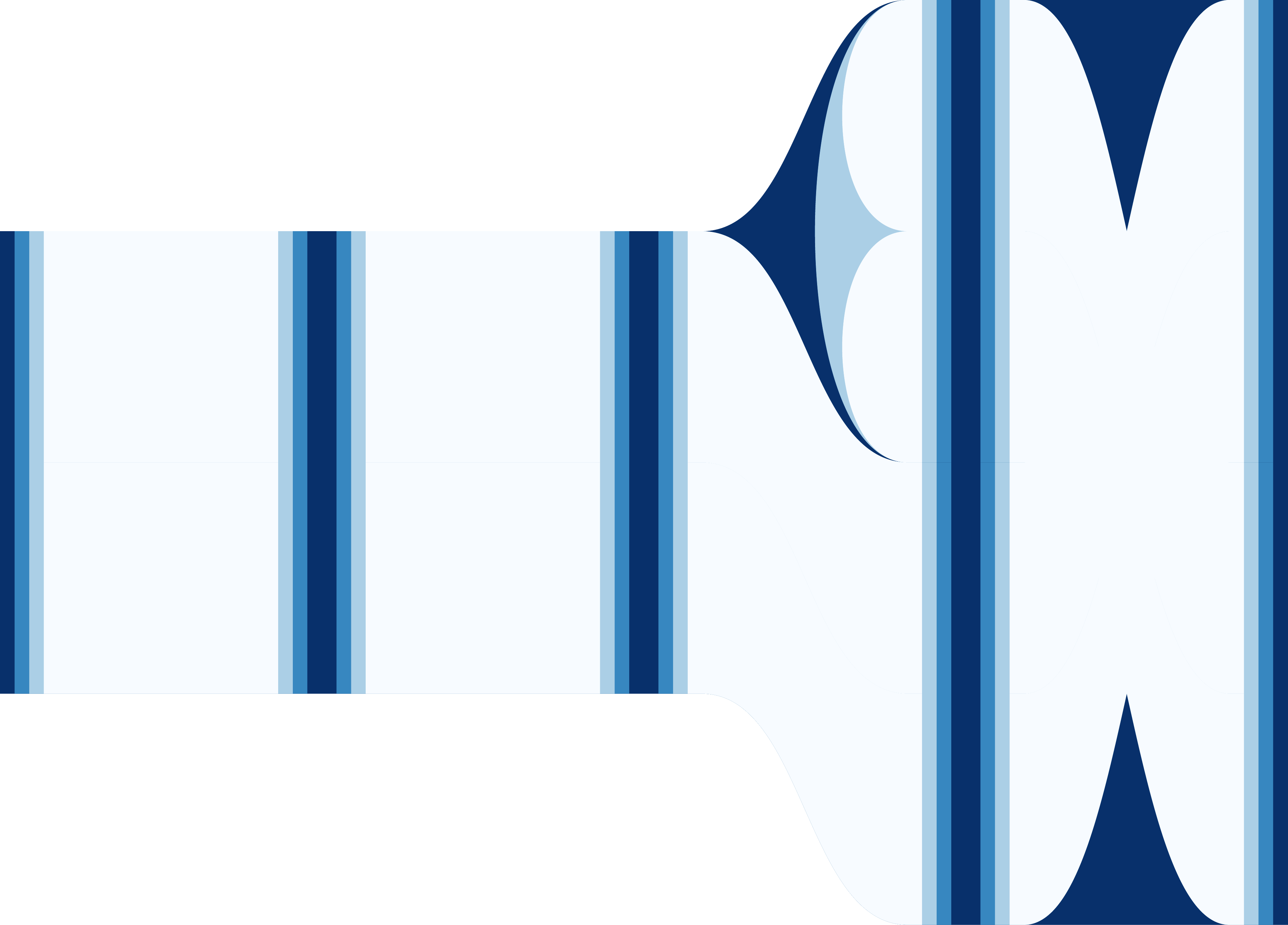}
        \put (14,7) {a) $y_m=0$}
        \end{overpic}
        }
    \vspace{0.2cm}
    \centering\scalebox{1}[1]{
        \begin{overpic}[height=3cm]{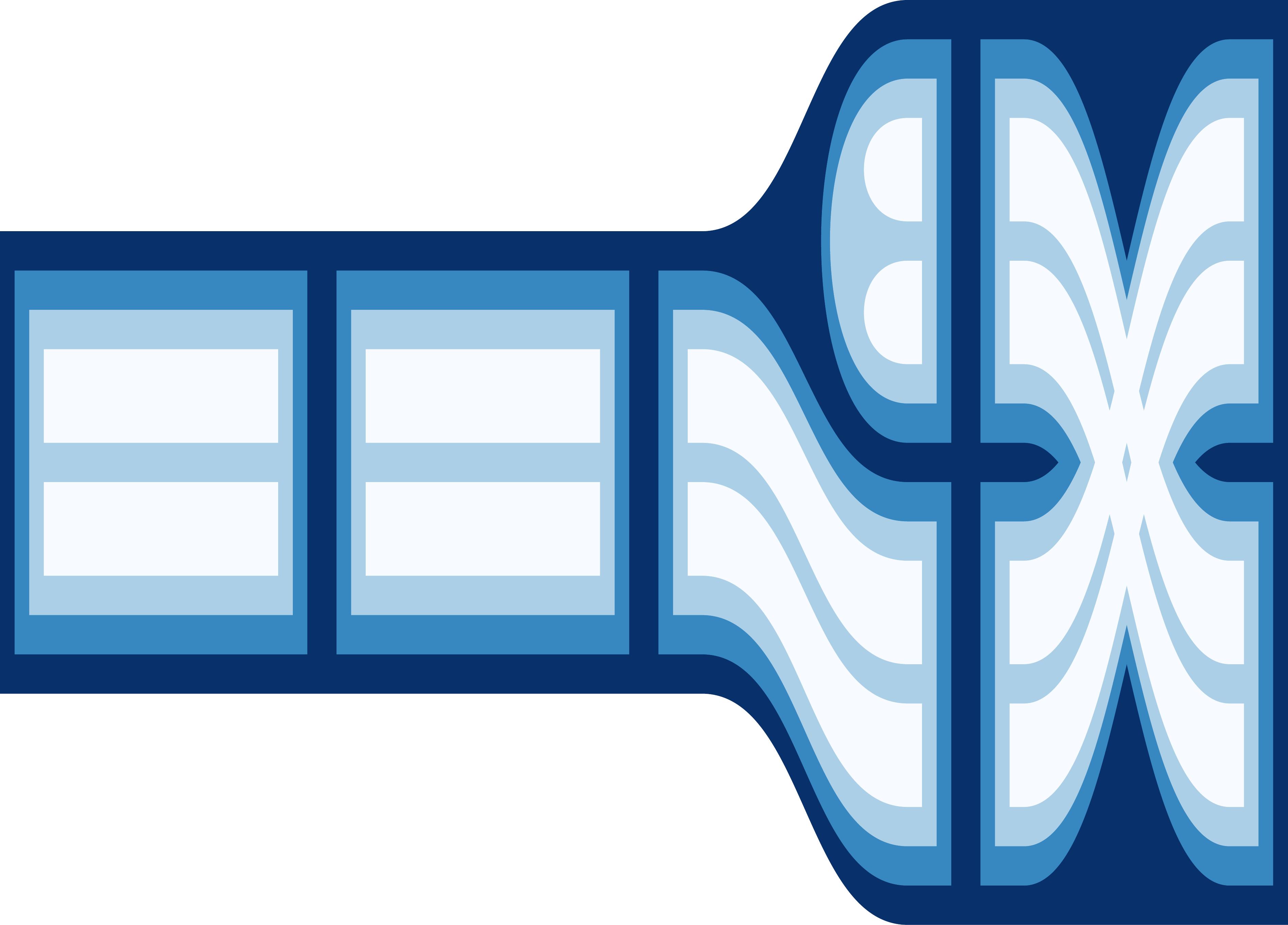}
        \put (14,7) {b) $y_m=10$}
        \end{overpic}
        }
    \vspace{0.2cm}
    \centering\scalebox{1}[1]{
        \begin{overpic}[height=3cm]{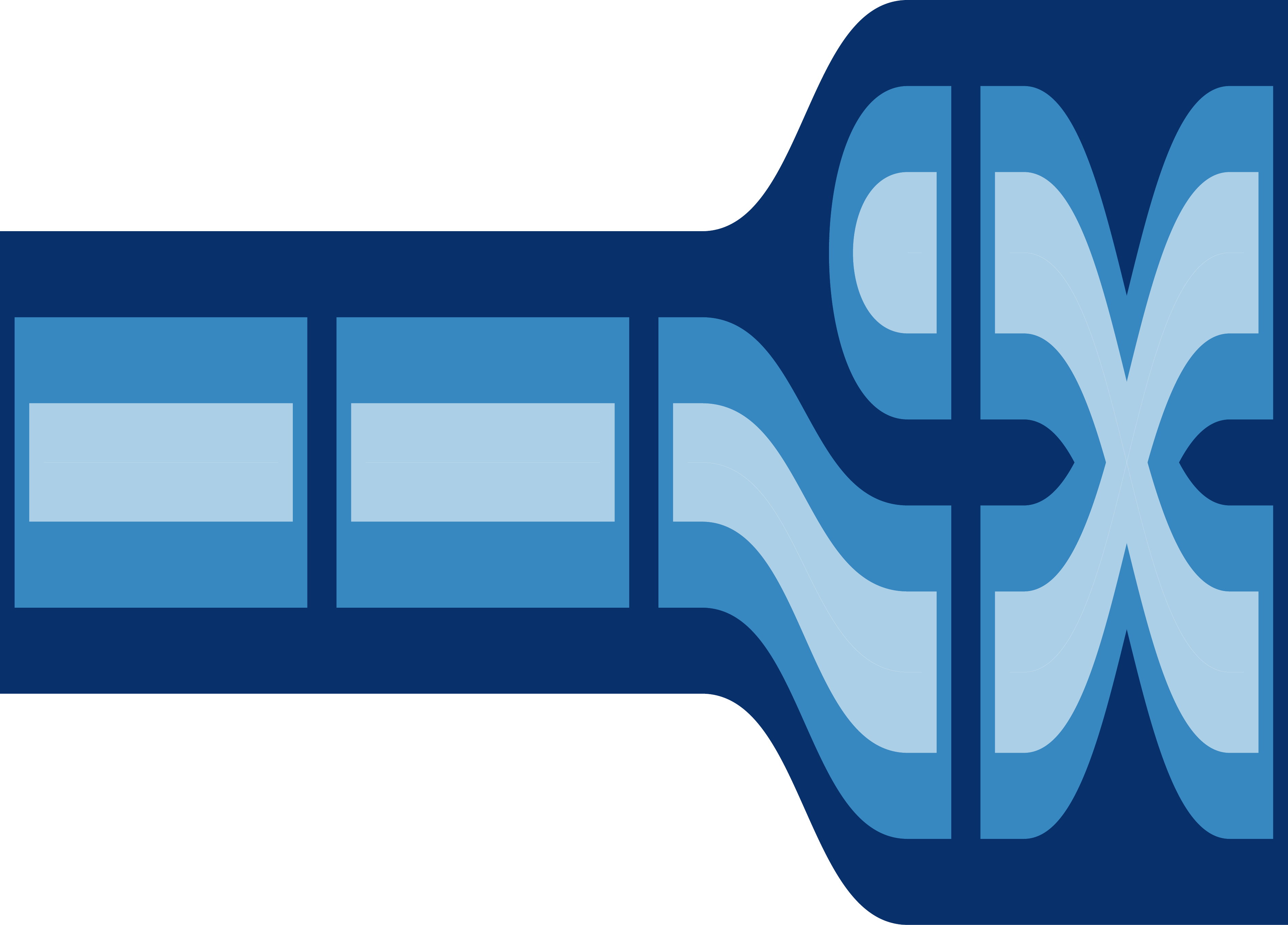}
        \put (14,7) {c) $y_m=20$}
        \end{overpic}
        }
    \vspace{0.2cm}
    \centering\scalebox{1}[1]{
        \begin{overpic}[height=3cm]{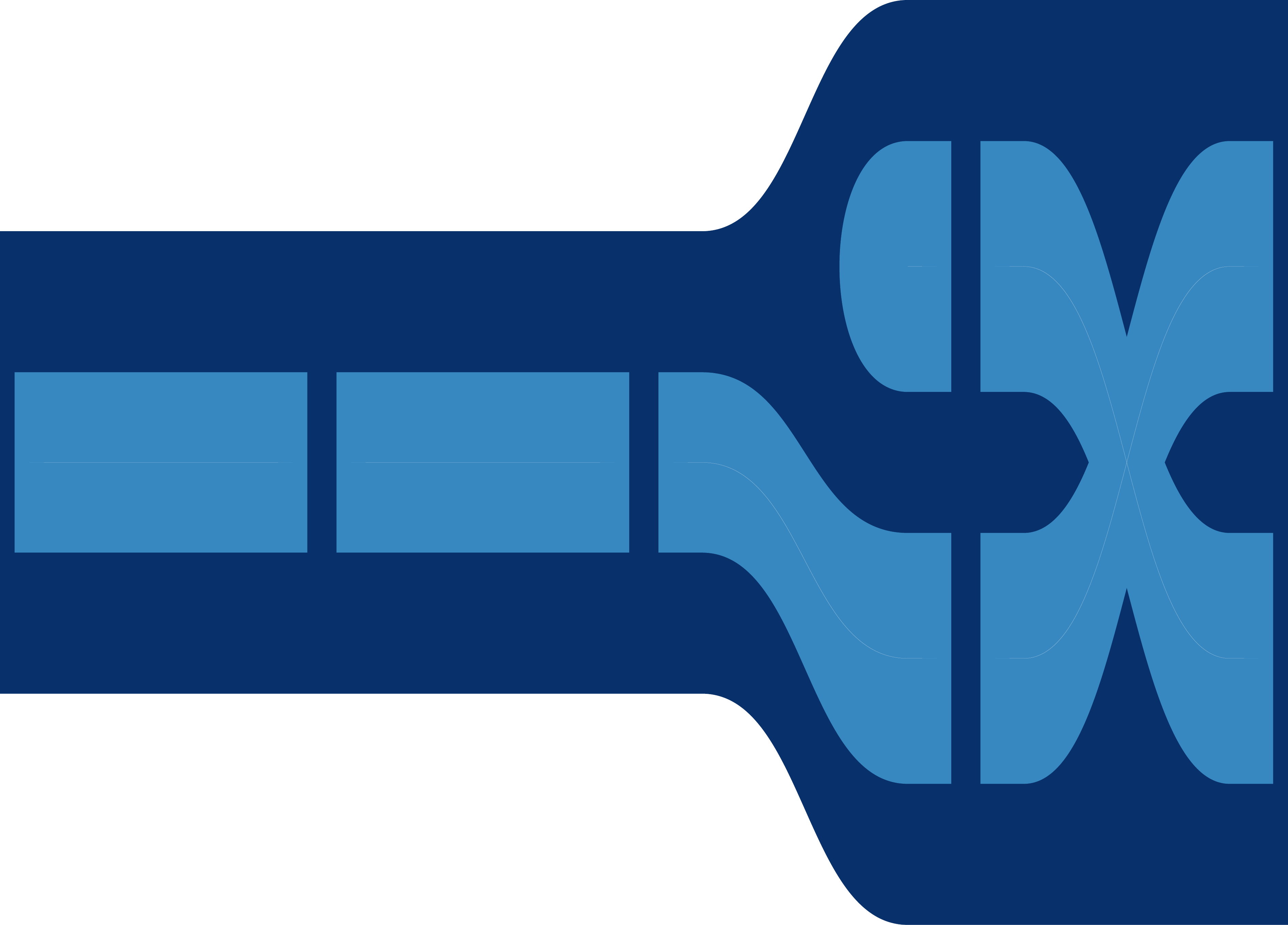}
        \put (14,7) {d) $y_m=30$}
        \end{overpic}
        }
    \caption{Y-Margin. For aggregated nodes (a), a y-margin can improve the perception of hierarchies (b). Increasing the y-margin (c,d) creates a form of semantic zoom, since smaller nodes disappear, as the y-margin gets larger.}
    \label{fig:zoom}
\end{figure}

In the same manner as an x-margin can be applied to streams to better communicate the hierarchical nesting  of nodes, a y-spacing can help to ensure that a node's representation is completely covered by its parent, hence improving the perception of the hierarchy. In particular for aggregated nodes, hierarchies are only visualized in the x-dimension where split. Y-spacing must be applied with care, as it distorts the represented values. In the following we will discuss two different approaches in detail and demonstrate their advantages and drawbacks.

Adding a y-padding to the visualization means increasing the value of every node which is not a leaf node. Let us consider a simple example, where the tree at a given point in time consists of a root node with one child. The child node has a value of 1, so the root node's aggregate is 1. If we draw the streams for that tree, the root will be completely occluded by its child. We could therefore increase the root's value to make it visible. While hierarchies would be better perceived that way, it also introduces an error because the root node is now represented larger than before. The deeper our hierarchy is, the more padding needs to be introduced. In many cases we want to see additional spacing between siblings, so the padding scales with both depth and the number of children. The most critical part is, that if a node changes its hierarchy level from one timestep to another but keeps the same value, that this value will be represented with different heights, because the applied y-padding depends on the depth of the node. This can have a major impact on the interpretation of values in the visualization. The benefit in perceiving hierarchical structures is in this case counterbalanced by reduced accuracy in value perception. However, for tasks in which the node values are not present or are not of primary importance, but the hierarchical nesting and structural changes matter, y-padding can be utilized as showcased in \autoref{fig:mesh1}, to increase the perception of hierarchies. We gave every leaf node a value of 1 and added a y-padding of 1 plus the number of children to every aggregate.

Instead of increasing a node's parent with y-padding, we can reduce a node's height directly by a fixed value. We call that value \textit{y-margin}. The error being introduced is very similar to that of y-padding, with the difference that the node's height will reach 0 as soon as the y-margin is larger than the node's value. This approach can be utilized as a form of semantic zoom, by increasing the y-margin step by step. As a result, nodes with small values will disappear and nodes of higher hierarchy levels become visible. This approach allows for data inspections independent of the hierarchical level, but only dependent on the node's values. We demonstrate the semantic zoom with different levels of y-margin in \autoref{fig:zoom}. In the same way the x-margin can be defined by different functions (\autoref{eq:fixed}-4), the y-margin can be defined to allow for task-dependent steering of the zoom functionality.

\subsection{Algorithm}


In order to generate a SplitStream, we need to traverse the given hierarchy at every point in time and calculate a position for every node of the tree. The nodes are then individually traversed through time and drawn based on their characteristics. We need to handle cases of nodes being created, being deleted, splitting, and merging, in addition to the connection of nodes by streams. When all streams are drawn, we can introduce splits to cut them open based on the defined margin function. In the following, we will describe our method in more detail.

If values are only defined in the leaf nodes of the data, or the hierarchy has no values defined at all, we need to compute missing values. We therefore recursively traverse every tree in a depth-first approach, until reaching a leaf node. If the leaf has no value defined, we set it to 1. We can then define every node's size by computing its aggregate. To make the hierarchies more visible in the final visualization, we can add a y-padding here.

In case the data does not specify positions of nodes, we can define them in such a way that all children are equally spread within their parent element. We initialize the position of the root element with 0 and iterate through all nodes of the tree except for leaves to define their \textit{spacing} attribute as follows:
\begin{equation}
    spacing_n \gets (\textit{n.size - n.aggregate}) / (\#children +1).
\end{equation}
It describes the space between all children of the given node $n$ in the resulting stream visualization. We can then compute the position of every child by:

\begin{algorithmic}
    \State
    \State $aggregate \gets 0$
    \ForAll{i:=1 \textbf{to} \#children}
        \State $child[i].pos \gets i \times spacing + aggregate$
        \State $aggregate \gets aggregate + child[i].size$
    \EndFor
    \State
\end{algorithmic}

When the position and size of every node are set, we can connect them to streams. We can identify all streams by looking for nodes which do not have a predecessor. Starting with these nodes, we follow all their succeeding nodes, draw a stream between the two nodes and repeat this operation in a recursive manner for all following nodes. The algorithm to draw all streams can be found in \autoref{appendix}A. In order to handle the special case in which a node moves into one of its ancestors, we need to check for every node which changes its parent, if that parent was an ancestor of this node in the previous timestep. If this is the case, we split the stream into two streams and mark them to draw their case-specific encoding at the beginning or the end.

Splits can be integrated into the drawing algorithm directly, or be applied as a post-processing step. The fact that splits can occur at any point in time and can have an arbitrary size, makes the analytical integration somewhat cumbersome. Instead, we apply splits after the drawing process by going through every stream and cutting it at positions where splits occur. For every split we can find the two nodes of the stream which are closest to the split (left and right) and remove a part of the stream equivalent to the respective margins of these nodes. For SplitStreams, we apply splits at every timestep, so that the nodes to the left and to the right of the split both refer to the same node from the timestep the split is applied to. This ensures that the hierarchy is represented in the same way for changes occurring before and after the timestep. 

\subsection{Implementation}
\label{sec:implementation}

We implemented SplitStreams as a JavaScript library based on D3~\cite{Bostock2011}. To showcase our results, we utilized standard web technologies (HTML5, CSS3) and Vue.js~\cite{lib:vuejs}. Streams are created as SVG paths following the outline of all nodes a stream contains. Hierarchical changes are displayed by Cubic Bezier curves with control points set to the center between both points in time, but horizontally to the nodes' y-position. This assures $G^1$ continuity at the common points. We utilize SVG clippaths to apply x-margins by removing parts of the paths after they have been drawn.

\begin{figure}[h]
    \begin{subfigure}{0.4\columnwidth}
        \hfill%
        \begin{tabular}{cc}\hline
        \# nodes & time [ms] \\ \hline
        14 & 2 \\
        347 & 13 \\
        2471 & 50 \\
        12214 & 247 \\
        47371 & 956 \\
        99548 & 2404
        \end{tabular}
    \end{subfigure}
    \hspace{0.7em}
    \begin{subfigure}{0.5\columnwidth}
    \vspace{0.6em}
        \includegraphics[width=\columnwidth]{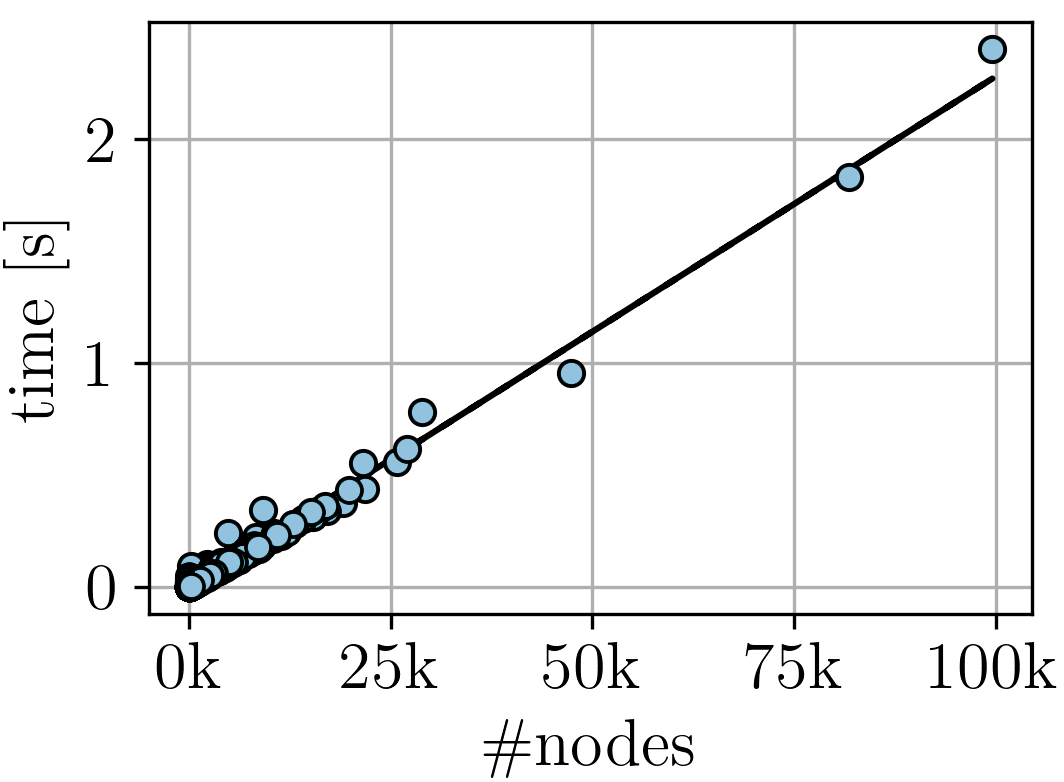}
    \end{subfigure}
    \caption{Algorithm performance for 1313 datasets. The benchmark was executed in Google Chrome v77 on a Windows 10 machine with an Intel Core i7-6700 CPU.}
    \label{fig:performance}
  \end{figure}

In order to evaluate the performance of our current implementation, we ran the algorithm for generating SplitStreams (\autoref{appendix}A) for a subset (1313 evolving hierarchies) of Vernier et al.~\cite{vernier2019quantitative}'s benchmark datasets ranging from small datasets up to approximately 100K nodes. The results are plotted in \autoref{fig:performance} (right) and the table on the left shows exact timings for selected representative datasets. The timings do not include rendering of the created SVG image as performance varies widely across browsers and devices. In the worst cases that we observed, for very complex topologies, the SVG rendering time was approximately the same as the generation itself. The complete implementation of our method, including many of the presented examples, can be found at \textit{https://github.com/cadanox/SplitStreams}.

%% file: s6_results.tex
\section{Use Cases}
\label{sec:results}

In order to showcase some of the results that can be achieved by applying our visual metaphor to real data, we selected a taxonomy of medical terms, as well as a public Github repository to demonstrate their evolution over time.

\subsection{MeSH Taxonomy}

Taxonomies are used in science to classify entities, like the phylogenetic tree which shows evolutionary relationships between species. The U.S. National Library of Medicine maintains a taxonomy of Medical Subject Headings (MeSH), currently including nearly 60000 medical terms in a hierarchically-organized vocabulary~\cite{mesh}. The taxonomy requires changes based on scientific progress and new insights gained, and versions are made available for the last \texttildelow 20 years. The visualization of changes over the years can help in finding inconsistencies and provide pointers to changes that should be made for the next revision. We selected two branches that show several interesting features of the hierarchy and visualized their evolution. Changes are specified between every two successive years. When there were no changes in the data for several consecutive years, we contracted them into one block.

\begin{figure*}
	\centering\scalebox{1}[0.9]{\includegraphics[width=\textwidth]{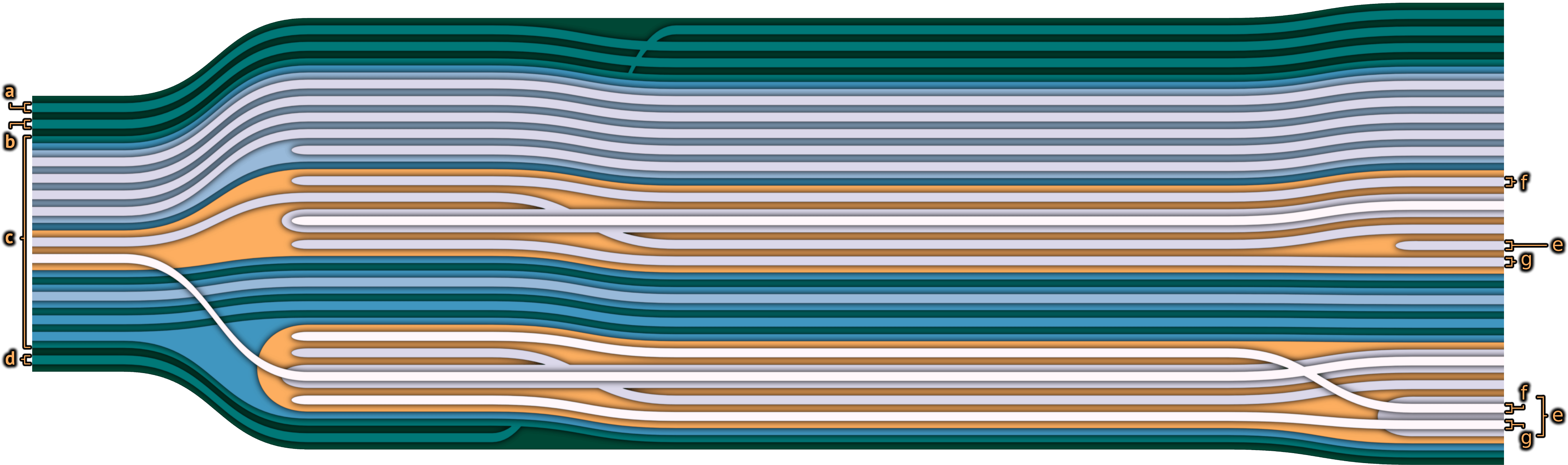}}
	\\\vspace{2mm}
	\centering\scalebox{1}[0.9]{\includegraphics[width=\textwidth]{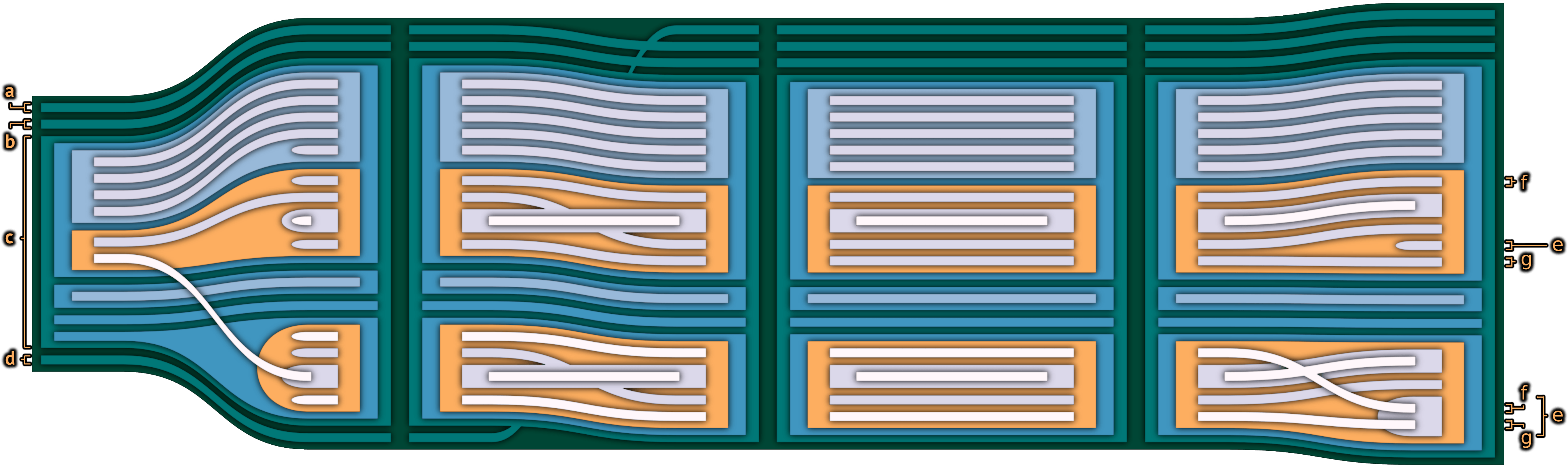}}
	\\\vspace{1mm}
	\centering{\includegraphics[width=\textwidth]{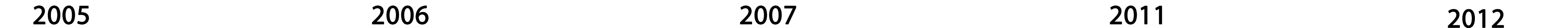}}
	\caption{MeSH Urinary Tract 2005-2012. Hierarchical changes of nodes being added, switching their sibling order, and moving into a new parent, are visible in both representations. SplitStreams enable the user to, e.g., count the number of timesteps shown (5) and count the number of main classes in the Urinary Tract (4). The additional space used for displaying lower level nodes eases the interaction task.}
	\label{fig:mesh1}
\end{figure*}

In \autoref{fig:mesh1}, we show the changes in medical terms for the Urinary Tract between 2005 and 2012 with the main branches Urethra (a), Ureter (b), Kidney (c), and Bladder (d). These classes are more clearly separated by SplitStreams than by a nested streamgraph. The two highlighted branches (orange) represent the Kidney Glomerus, once nested inside Nephrons and once inside the Kidney Cortex. In 2006, several nodes were added and the same structural changes were applied to both branches throughout the years. In 2012, the new class Glomerular Filtration Barrier (e) is introduced to the Kidney Glomerus in both Kidney Cortex and Nephrons. However, the two classes Glomerular Basement Membrane (f) and Podocytes (g) are only defined as children of the barrier in Kidney Cortex. This difference is likely to be a mistake in the data and should be fixed in the next revision of the taxonomy. The addition of splits helps to highlight the occurrence of hierarchical changes and to identify the depth of involved nodes.

\begin{figure*}
	\centering{\includegraphics[width=\textwidth]{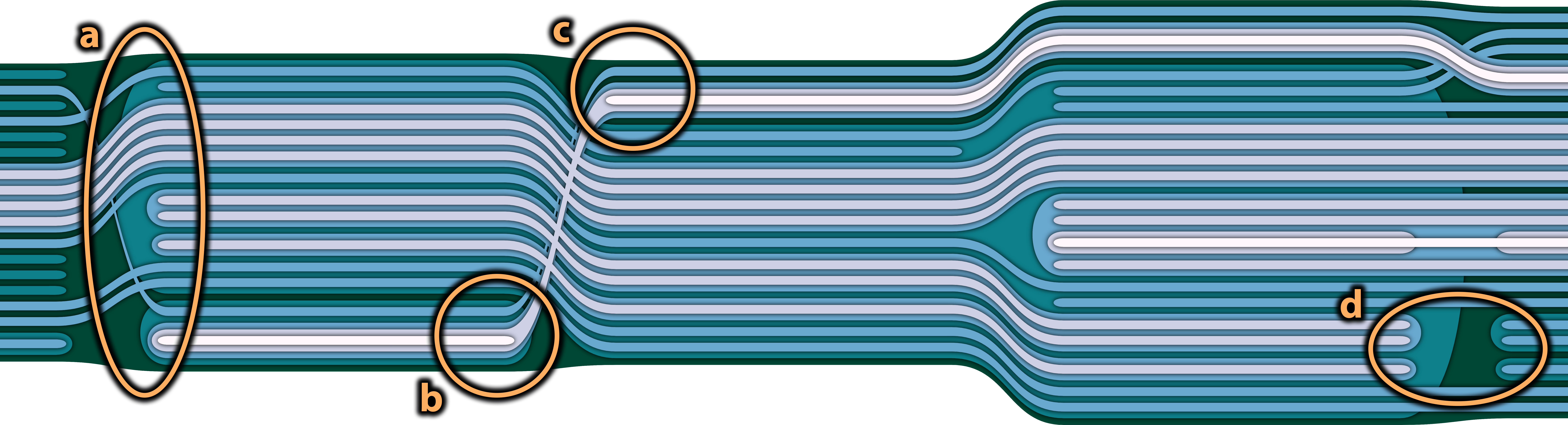}}
	\\\vspace{2mm}
	\centering{\includegraphics[width=\textwidth]{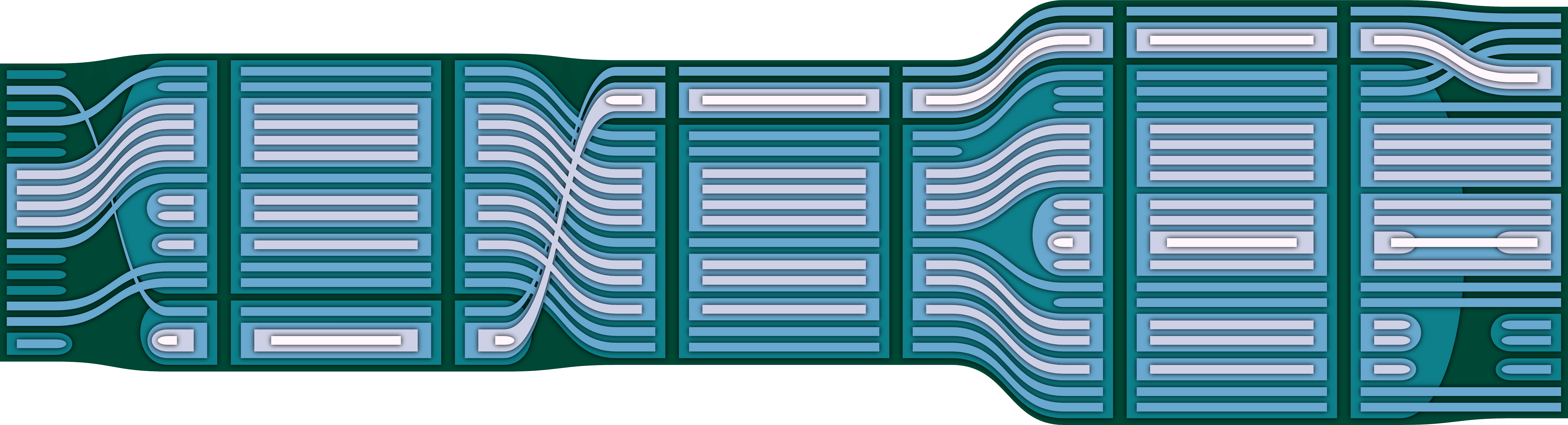}}
	\\\vspace{1mm}
	\centering{\includegraphics[width=\textwidth]{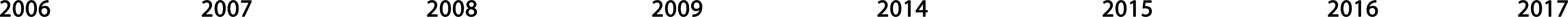}}
	\caption{MeSH Digestive Physiology 2006-2017. a) Addition of Processes and Phenomena to the root node. b) Removal of the Phenomena class. c) Individual Phenomena move into the root node. d) Some changes of class IDs are not listed in the data.}
	\label{fig:mesh2}
\end{figure*}

The evolution of medical terms for the Digestive Physiology between 2006 and 2017 can be seen in \autoref{fig:mesh2}. In 2007, the two classes Processes and Phenomena are introduced on the root level (a). The hierarchy stayed consistent for a year, until Phenomena were removed again (b) and their children moved into the root node (c). In 2017, the Processes class is removed as well and all its children move to the root node while receiving new IDs (d). We can see that a few nodes were removed and added, instead of appearing as a continuous stream. This finding can indicate missing ID changes in the data, or an error in the data processing pipeline.

While most changes can be detected in the nested streamgraph representation, it becomes increasingly harder to understand the hierarchical structure and the depth of changing nodes. SplitStreams help us to map the continuous streams to their discrete timescale, highlight positions of hierarchical changes, identify the depth of individual nodes, and provide us with a structured representation of the underlying hierarchy. We can get a more general intuition of the depth layers in which changes occur and thereby judge their impact on the data.

\subsection{Leaflet Github}

\begin{figure}
	\centering\includegraphics[width=\columnwidth]{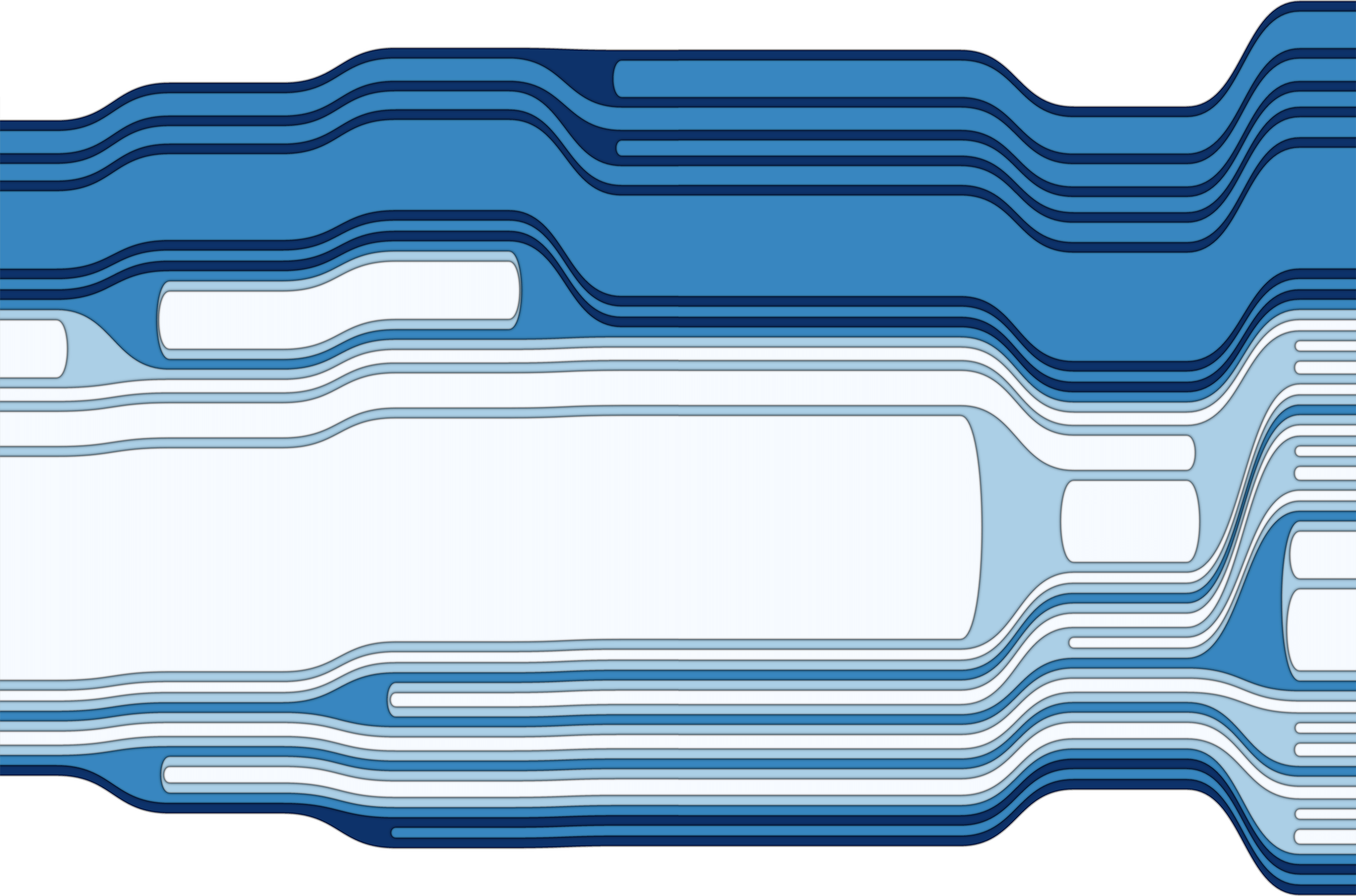}
	\\\vspace{2mm}
	\centering\includegraphics[width=\columnwidth]{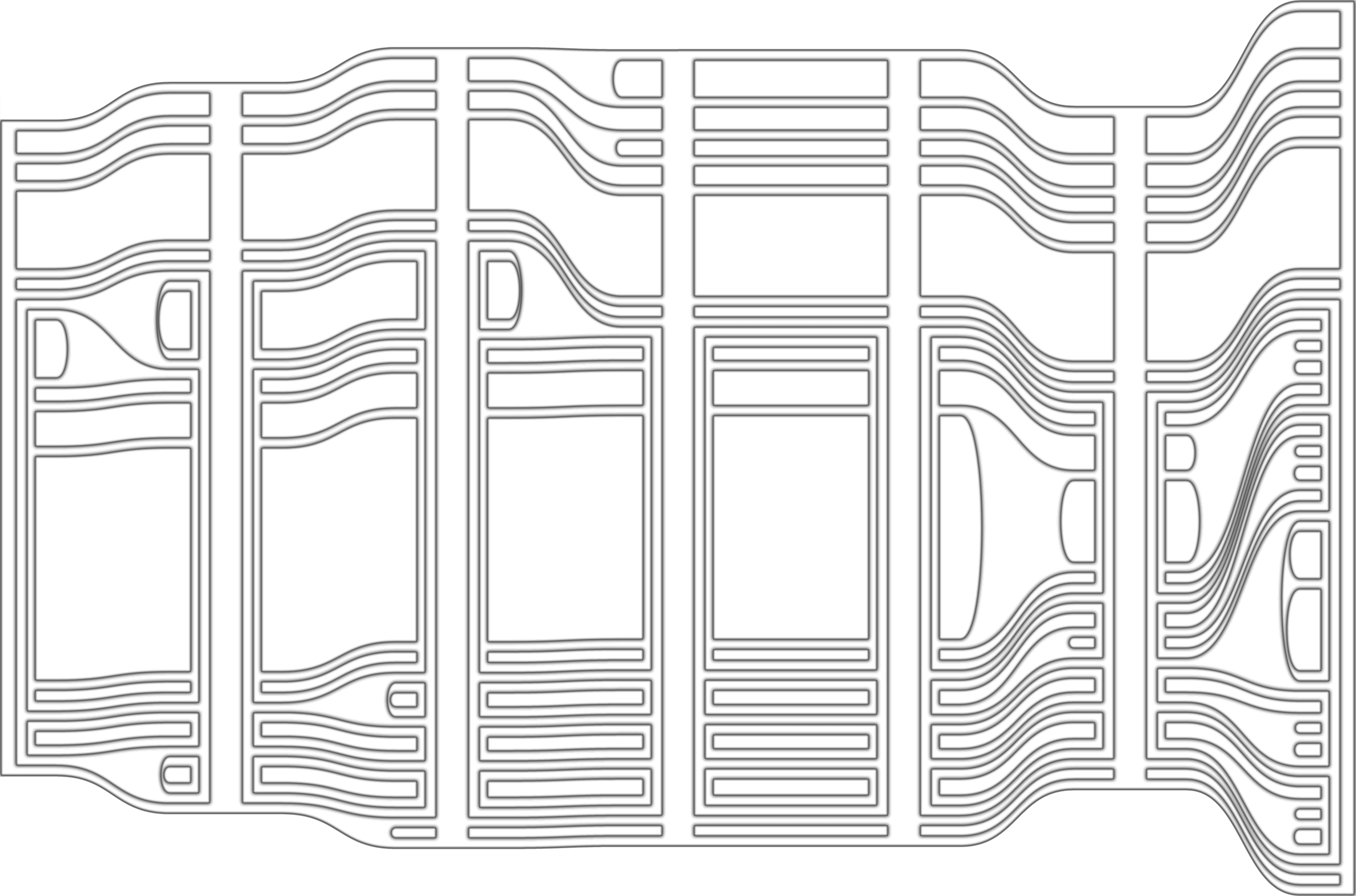}
	\caption{Dear ImGui Github Repository (7 monthly revisions). While nested streamgraphs (top) utilize color to convey the hierarchical structure of the data, SplitStreams represent it in the form of clear shapes. The hierarchy is even visible, when only drawing the outline of each node.}
	\label{fig:github}
\end{figure}

Vernier et al.~\cite{vernier2019quantitative} collected 2720 datasets showing the evolution of values over time, roughly half of which feature a hierarchy. While the authors utilized the data to benchmark different treemap layouts with regards to their stability, we can visualize the very same data in our static, time-dependent hierarchy representation.
Dear~ImGui~\cite{imgui} is a graphical user interface library for C++. In \autoref{fig:github} we show the evolution of 7 monthly revisions of the ImGui Github repository, once as nested streamgraph and once as SplitStream. The folder structure builds the hierarchy at each point in time and each file's value is determined by the number of lines of code the file contains. While the changes of values and hierarchical changes (addition, deletion of files) are visible in the nested streamgraph, it can be difficult to understand the hierarchical structure given in each point in time. While nested streamgraphs utilize color to represent a node's depth, SplitStreams make use of clear shapes and show depth via containment. A better understanding of the hierarchy eases the task of correlating changes in the hierarchy to files in the repository. Being able to count the number of elements in each node (e.g., 5 in the first revision) gives us a better impression of the distribution of files and the size of higher-level folders. Being able to see a clear separation at every timestep further provides us with a better intuition for the time scale being used.
All benchmark datasets from Vernier et al.~\cite{vernier2019quantitative} can be investigated through the exemplary implementation of our method in the supplemental material.

%% file: s7_evaluation.tex
\section{Evaluation}
\label{sec:evaluation}

We conducted a controlled user study to investigate and compare our visualization technique with existing methods.
In this section, we briefly explain the experimental design and results.

\subsection{Motivation}

Since SplitStreams are an extension of the nested streamgraph (NSG) approach~\cite{lukasczyk2017nested, wittenhagen2016chronicler}, we want to see if the introduced design changes improve the users' performance in understanding the hierarchical structure at specific points in time. Treemaps are a well established technique for visualizing hierarchies, so that we can use them as a baseline for how well hierarchies can be perceived in a visualization. Temporal Treemaps \cite{kopp2019temporal} utilize cushions to emphasize hierarchical structure in NSG and can therefore be seen as the logical competitor to our approach. Since cushions can be applied to all our tested visualizations, and the introduction of yet another visual encoding would increase the number of independent variables and thereby decrease the statistical power of the analysis and results, we do not consider them for this comparison.

\subsection{Hypotheses and Goals}

Based on the main tasks this visualization is meant to solve, we considered the performance in understanding hierarchical structure at a given point in time, understanding changes in hierarchy, and comparing node values over time.


Given that our method uses the same shape-based design as \tree{} to visualize node containment, we hypothesized that users would have a better performance using \secs{} compared to \stream{} in understanding hierarchies at a given point in time.
Since hierarchical changes over time are equally represented in \stream{} and \secs{}, we expected users' performance to be significantly better than in the \tree{} visualization, where changes are not explicitly displayed.
Finally, we tested the hypothesis that both \secs{} and \stream{} would be superior to the \tree{} design for the task of comparing node values, because the streams help to identify the areas of interest.

\subsection{Experiment Design and Task}

To test our hypotheses, we designed an experiment in which the participants would view images of visualization techniques and answer one question per image.
We defined one independent variable for this experiment, \textit{visualization type}, with three levels: \secs{}, \stream{}, and \tree{}.
In order to not overwhelm the participants and avoid learning effects, the study followed a between-subject design, where each participant would complete exactly one condition.
Each participant completed 14 basic analysis tasks each using a unique visualization image. To avoid potential confounds, all conditions had the same tasks, featuring the same datasets, but showing different images based on the condition created by the system described in \autoref{sec:implementation}.

The questions were based on a file system scenario with folders and sub-folders and designed to cover three different analysis tasks. The first task type focused on understanding the hierarchy at a certain point in time (e.g., count the number of siblings or ancestors a certain node has at a certain timestep). The second task type focused on understanding changes in hierarchies over time (e.g., count the number of times a node moved). The last group focused on node-value comparisons over time (e.g., how many nodes shrank during a given period of time). We had 13 questions with a distribution of 5-5-3 for the different task types, and one simple attention check question that was excluded from the analysis.
The order of trials was randomized to avoid potential confounds.
To reduce task complexity, trials included no more than 20 streams and 5 timesteps in one image. Additionally, the number of hierarchical changes was kept low to avoid frustration particularly in the \tree{} condition. All questions for all conditions can be found in the supplemental material.

Participants provided answers via numeric text entry and could make multiple attempts. Each trial participants were required to enter the correct answer or reach a 5-minute time limit before continuing to the next question.
Correctly guessing the correct answer without inspecting the visualization was highly unlikely due to the unbounded nature of the textbox input.

Metrics used to assess user performance included: \err{}, \attemptTime{}, \attempts{}, and \tme{}, recorded for each trial. 
Error was calculated as the absolute difference between the answer given in the first attempt and the correct answer. 
To record and report these measures in accordance to our study goals, we calculated the average of each metric based on the question type.

\subsection{Participants and Procedure}
The evaluation was conducted as an online user study using Amazon Mechanical Turk (AMT). A total of 120 participants completed the study, of which 102 passed the attention check and were included for the analysis. 
We ended up having 34, 35, and 33 participants per \tree{}, \stream{}, and \secs{} conditions.
Participants first reviewed a set of instructions based on their assigned condition, followed by 3 example trials and 3 practice trials. 
After each example and practice trial, they were shown feedback with the correct answer and a visual explanation for how the answer was achieved.
Next, participants completed the main set of questions to provide the results for analysis.
In order to avoid any learning effects, participants were not given feedback with the correct answers after answering during the main tasks.



\subsection{Results}

\begin{table*}[t] 
    \centering
    \setlength{\tabcolsep}{13pt}
    \renewcommand{\arraystretch}{1.5}
    \begin{tabular}{|l|c||rr|}
         \multicolumn{2}{c}{\textbf{Measure}} &
         \multicolumn{1}{c}{\textbf{Understanding hierarchies}} &
         \multicolumn{1}{c}{\textbf{Understanding changes of hierarchies over time}} \\
        \hline
         \multirow{2}{*}{\textbf{Average \# of Attempts}} & Main Effect &
         $ \chi^2(2) = 24.39$, ($p < 0.001$) * &  
         $ \chi^2(2) = 14.65$, ($p < 0.001$) * \\ \cline{2-4}
         & \multirow{2}{*}{Posthoc Test} & \textbf{\secs{}} vs. \stream{} $(p < 0.001)$ * &  \textbf{\secs{}} vs. \stream{} $(p < 0.01)$ *   \\
         &&  \textbf{\tree{}} vs. \stream{} $(p < 0.001)$ * &  \textbf{\tree{}} vs. \stream{} $(p < 0.01)$ *  \\
        \hline
         
         \multirow{1}{*}{\textbf{Total Time per Task}} & Main Effect &
         $ \chi^2(2) = 7.80$, ($p < 0.05$) * &  
         $ \chi^2(2) = 10.58$, ($p < 0.01$) * \\ \cline{2-4}
         & \multirow{2}{*}{Posthoc Test} & \textbf{\secs{}} vs. \stream{} $(p < 0.05)$ * &  \textbf{\secs{}} vs. \tree{} $(p < 0.01)$ *   \\
         && & \textbf{\stream{}} vs. \tree{} $(p < 0.05)$ * \\
        \hline
         
         \multirow{2}{*}{\textbf{Error of First Attempt}} & Main Effect &
         $ \chi^2(2) = 26.80$, ($p < 0.001$) * &  
         $ \chi^2(2) = 14.60$, ($p < 0.001$) * \\ \cline{2-4}
         & \multirow{2}{*}{Posthoc Test} & \textbf{\secs{}} vs. \stream{} $(p < 0.001)$ * &  \textbf{\secs{}} vs. \stream{} $(p < 0.01)$ *  \\
         &&  \textbf{\tree{}} vs. \stream{} $(p < 0.001)$ * &  \textbf{\tree{}} vs. \stream{} $(p <0.01)$ *  \\
        \hline
         
         \multirow{2}{*}{\textbf{Time for First Attempt}} & Main Effect &
         $ \chi^2(2) = 4.75$, ($p = 0.09$) &  
         $ \chi^2(2) = 15.32$, ($p < 0.001$) * \\ \cline{2-4}
         & \multirow{2}{*}{Posthoc Test} &  & \textbf{\stream{}} vs. \tree{} $(p < 0.001)$ * \\
         && &  \textbf{\secs{}} vs. \tree{} $(p <0.05)$ * \\ \toprule
         
    \end{tabular}
    \caption{The study results for the task types of \textit{understanding hierarchy} and \textit{understanding changes of hierarchies over time} using a non-parametric Kruskal-Wallis test and a Wilcoxon post-hoc test. We only show the pairwise comparison in the posthoc test if the two conditions had a significant difference. Significant p-values are marked with an asterisk, and condition names with superior performance are shown in bold. Note that the results from the \textit{node-value comparison} task are \textbf{not included} in this table for simplicity.}
    \label{table:studyResults}
\end{table*}

We compared the results from \secs{}, \stream{}, and \tree{} conditions to understand their differences based on task types.
Due to the data not being normally distributed, we used the Kruskal-Wallis non-parametric test to evaluate the main effect differences among the study conditions and a Wilcoxin post-hoc test for pairwise comparison.
The study results are shown in \autoref{table:studyResults}.
Due to space restrictions, we do not report the test results from \textit{node-value comparison} tasks in this table as we only observed one significant effect among all four measures.

\begin{figure}[!t]
\captionsetup{justification=centering}
    \begin{subfigure}{0.49\columnwidth}
        \centering{\includegraphics[width=\columnwidth]{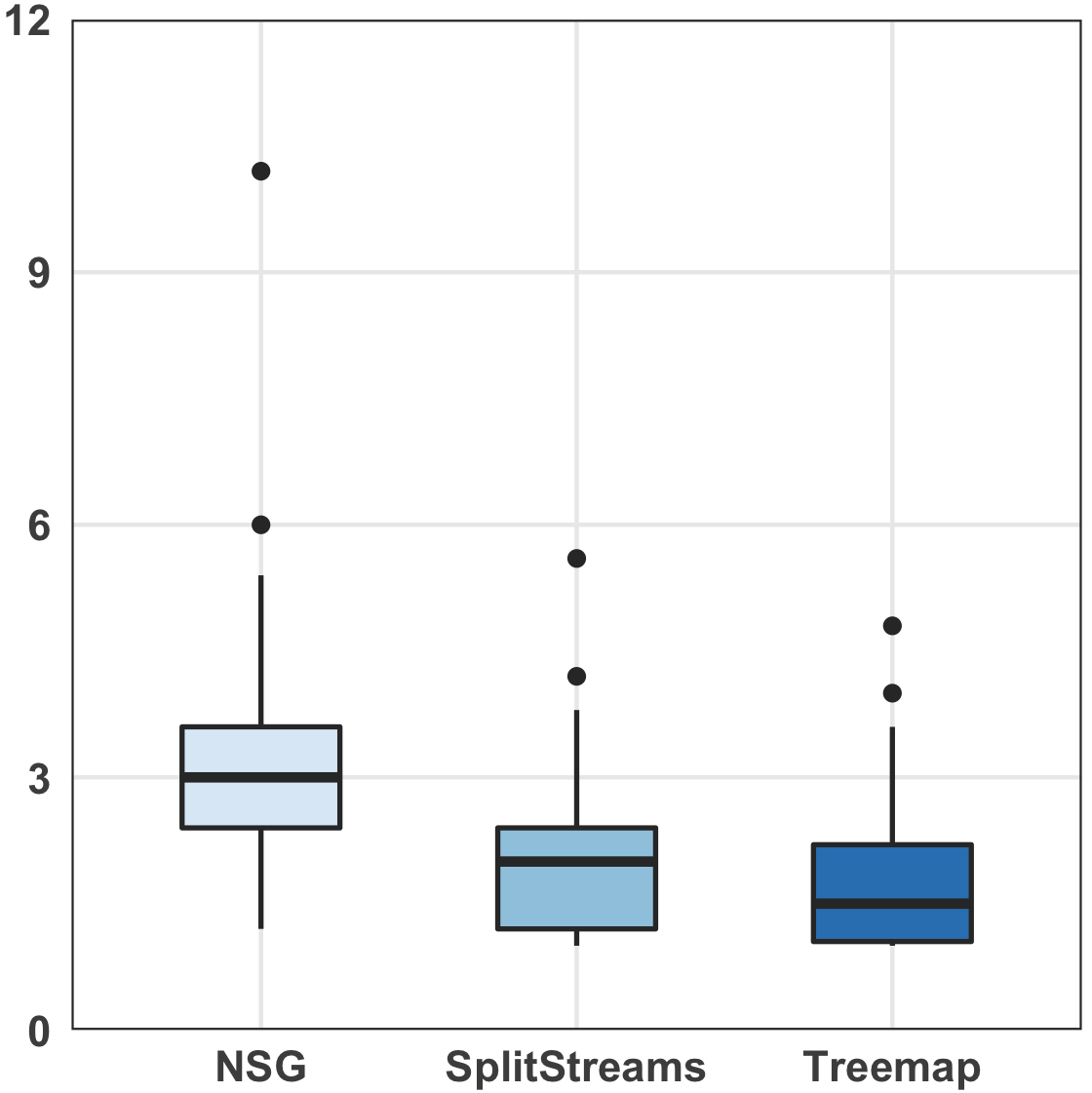}}
        \caption{Average Number of Attempts in \textit{Understanding Hierarchies}}
        \label{fig:a1}
    \end{subfigure}
    \begin{subfigure}{0.49\columnwidth}
        \centering{\includegraphics[width=\columnwidth]{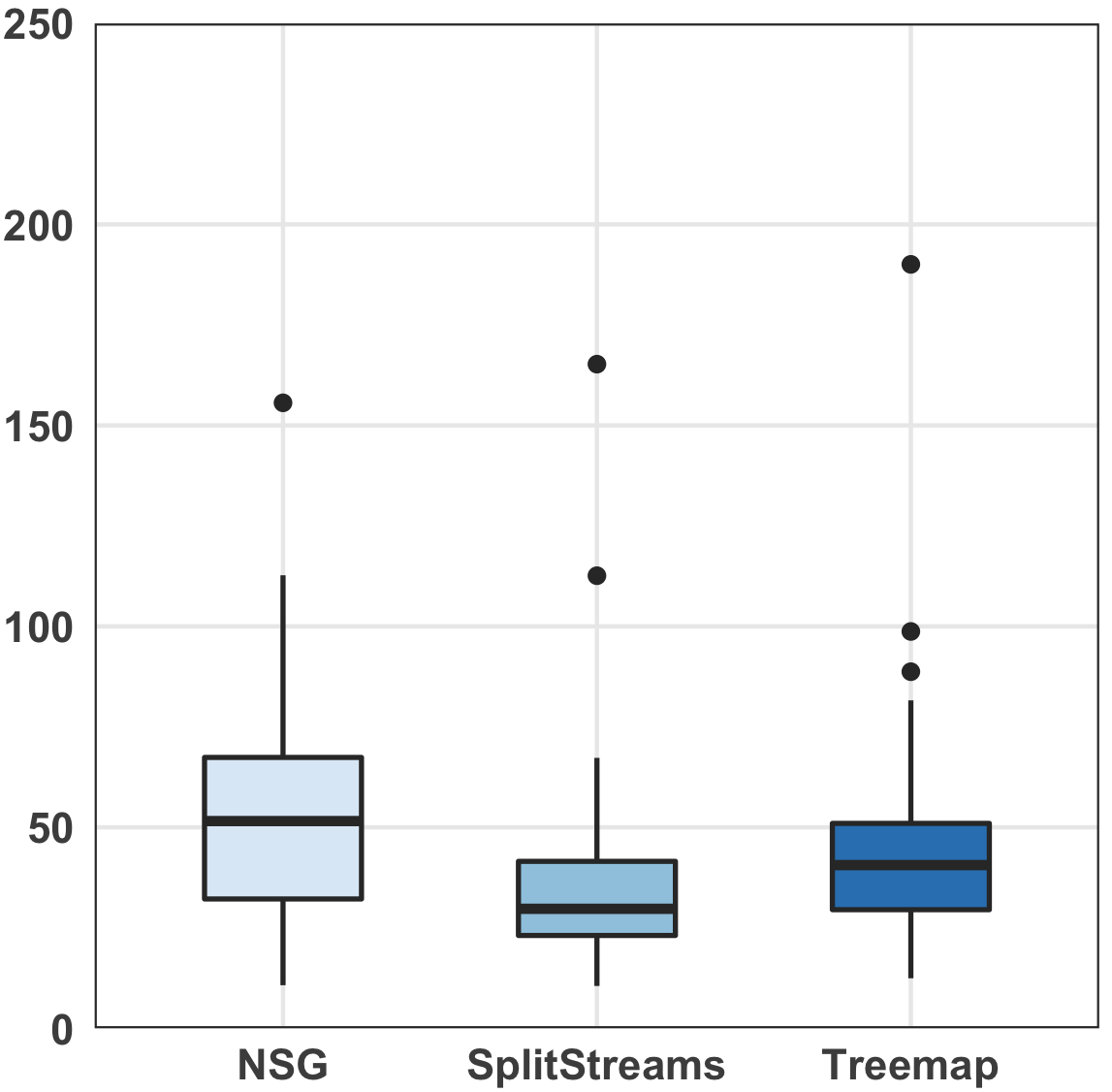}}
        \caption{Total Time (s) per task in \textit{Understanding Hierarchies}}
        \label{fig:a2}
    \end{subfigure}
    \begin{subfigure}{0.49\columnwidth}
        \centering{\includegraphics[width=\columnwidth]{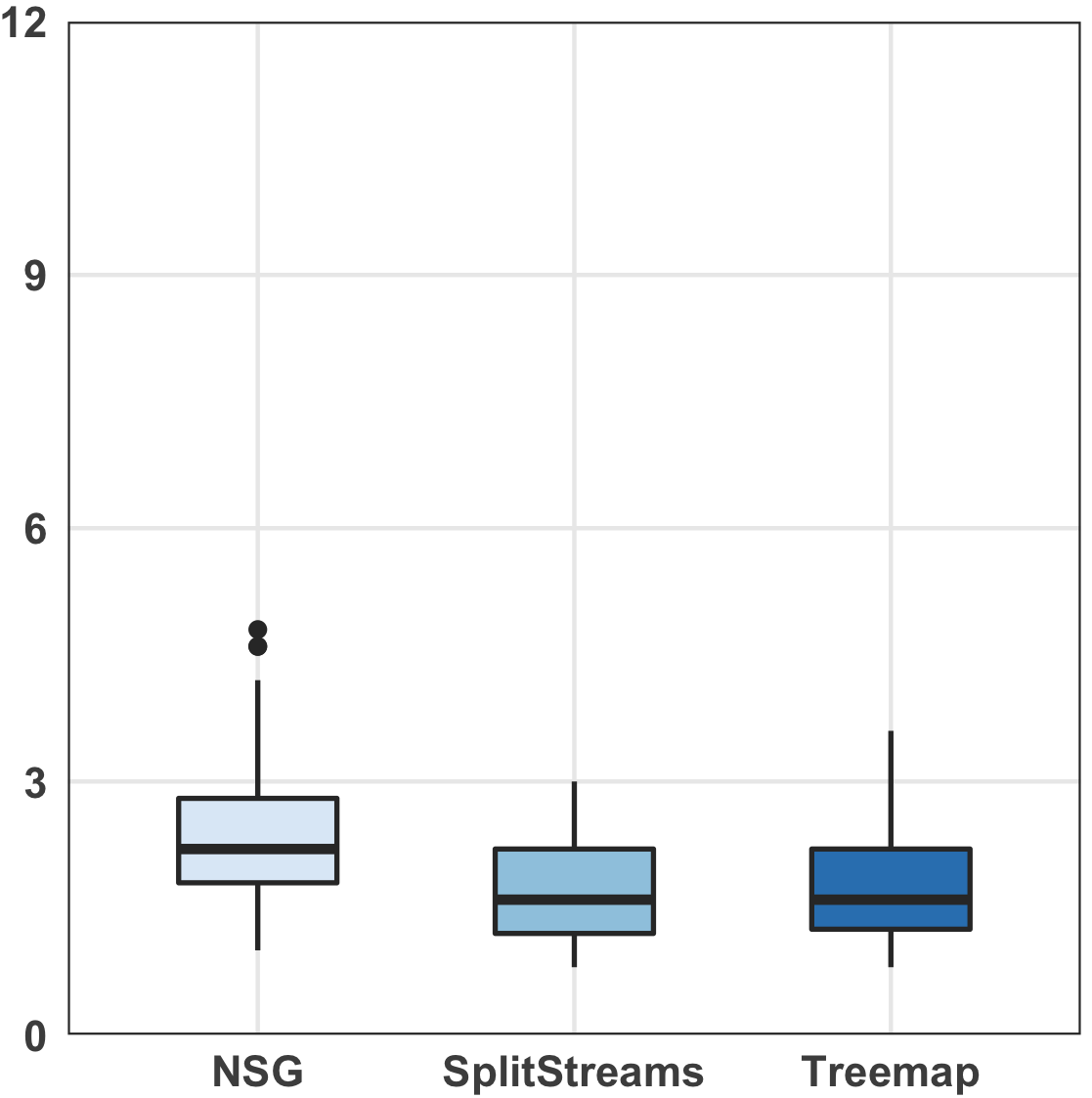}}
        \caption{Average Number of Attempts in \textit{Hierarchy over Time}}
        \label{fig:b1}
    \end{subfigure}
    \begin{subfigure}{0.49\columnwidth}
        \centering{\includegraphics[width=\columnwidth]{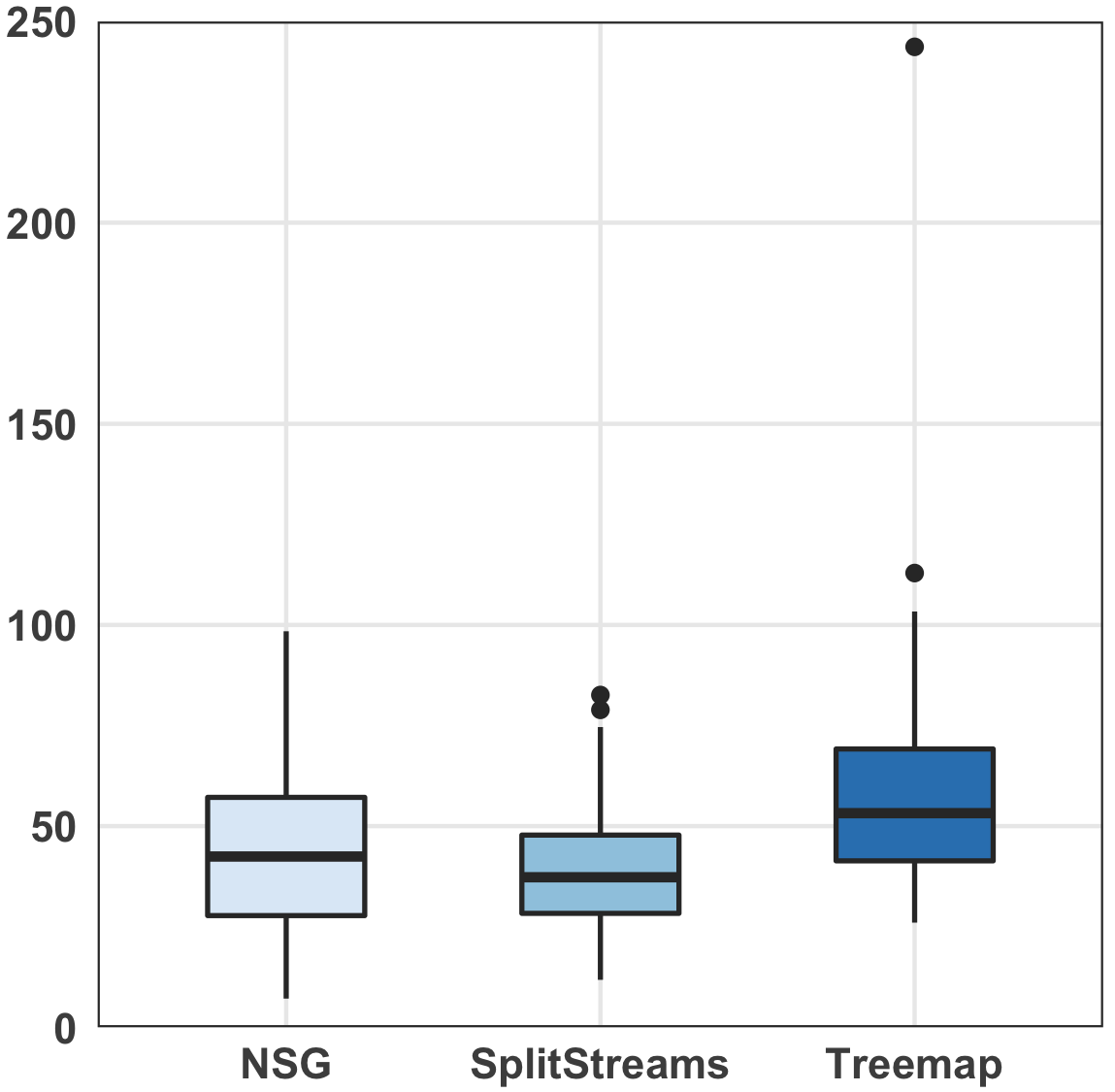}}
        \caption{Total Time (s) per task in \textit{Hierarchy over Time}}
        \label{fig:b2}
    \end{subfigure}
    \caption{Distribution of the results for the average number of attempts and the total time per task in seconds. 
    }
    \label{fig:userStudyCharts}
\end{figure}

\autoref{fig:a1} and \autoref{fig:a2} show the average \attempts{} and \tme{} for the task type \textit{understanding hierarchies}. 
The results show that the participants from both \secs{} and \tree{} conditions had a significantly lower \attempts{} compared to the \stream{} condition.
Also, participants were significantly faster in answering this type of question with \secs{} than with the \stream{} visualization.
These two observations align with our first hypothesis that our approach improves user performance compared to \stream{} by using the same shape-based design as \tree{}.

For the task focusing on \textit{understanding the changes of hierarchies over time}, our results demonstrate that \err{} and \attempts{} were significantly higher in the \stream{} condition than in \secs{} and \tree{} (\autoref{fig:b1}).
However, users of \tree{} were significantly slower in answering such questions than in the other two conditions (\autoref{fig:b2}).
Combining these findings, we can conclude that users from both \secs{} and \tree{} conditions were not significantly different in getting the correct answers, but people who used our technique could reach the conclusion faster.

Another interesting finding shows that users of \stream{} spent significantly less time on their first attempt compared to users of \tree{}, while also having a significantly higher \err{}.
This could mean that users of \stream{} were more confident, hence providing their answer faster, but were wrong more often. 
One possible interpretation of this result could be that stream-based techniques put less cognitive load on users than \tree{}.

Our evaluation of node-value comparison tasks showed a significant difference for \err{} ($\chi^2(2) = 9.10$, $p < 0.05$).
The post-hoc test showed that users had significantly lower errors with \secs{} than with \stream{} ($p < 0.01$). 
This suggests that our approach helps users understand changes of values over time better than with \stream{} at the first glance. 
We do not have enough evidence to support or reject our hypothesis that \tree{} users would perform worse than stream-based approaches in this task.

Overall, the evaluation was unable to detect major drawbacks in \secs{} as compared to the tested alternatives.
Our findings provide evidence to support our hypotheses that by using \secs{}, we can utilize the benefits of both \stream{} and \tree{} while avoiding some of their individual shortcomings.

%% file: s8_discussion.tex
\section{Discussion}
\label{sec:discussion}

The technique presented in this paper can be used to generate one-dimensional treemaps over time, nested streamgraphs, as well as new representations introduced by x- and y-margins. 

The results of our study demonstrate that \secs{} provide a similar performance to \tree{} in tasks involving understanding hierarchical structures in an isolated time period. 
This shows the approach can convey a similar level of detail in terms of hierarchical structure while also providing additional information on changes in the data.
With both stream-based approaches providing statistically-significantly faster user responses and better performance than \tree{} in tasks for understanding changes over time, \secs{} represents a general-purpose technique with good performance in the tested use cases.

When it comes to scalability, the number of nodes and timesteps that can be visualized by SplitStreams without introducing clutter are in the same order of magnitude as in treemaps and nested streamgraphs.
Based on our experience and using current screen resolutions, a static hierarchy might display up to several hundred distinguishable nodes, but this capacity is highly diminished with the introduction of hierarchical changes. While value changes, additions, and deletions of nodes can occur in higher quantities, the visual overlap introduced by node movements can greatly affect the user's capabilities of reading the visualization.
Although we try to handle visual clutter through the introduction of a y-margin as a form of a semantic zoom, the exploration of large datasets (e.g., the visualization of the whole MeSH taxonomy in \autoref{sec:results}) requires additional mechanisms to ensure scalability.
While SplitStreams and treemaps allow for a better hierarchical perception in deep hierarchies, their introduced margins require more space along the time axis than nested streamgraphs. To be precise, a deeper hierarchy and larger margin definition requires more distance between individual timesteps (\autoref{equ:HCR}).
On the other hand, the margin space can not only be exploited to enhance and highlight the hierarchical structure with, e.g., halos or drop shadows, but also eases the selection of a node of interest in an interactive environment.
The discontinuity along the temporal axis introduced by \secs{} had no detectable negative impact on user performance in our study, but might affect the aesthetic appeal or visual complexity of the representation.

While we currently apply splits to all streams at every single point in time, we could apply margins to a subset of streams and timepoints.
Such an approach could, for instance, emphasize hierarchical changes by only applying margins to streams for which a hierarchical change happens.
Conversely, since nested streams do a good job at representing hierarchical changes, it would be possible to selectively emphasize the hierarchy when no changes occur. 
We believe that such a selective application of our technique could allow for an enhanced depiction of features of interest.

We introduced three different functions to apply x-margins at points of splits. 
While these functions can be defined in many ways and adjusted based on specific user tasks, their advantages and disadvantages deserve further study.
The same is true for y-spacing in particular because it affects node value perception.

Existing work has shown that the order of streams can be optimized to reduce the number of stream crossings~\cite{kopp2019temporal}. 
However, the proposed strategy is limited to data where hierarchical changes occur along siblings.
Our technique would benefit from an adapted ordering algorithm to consider hierarchical changes of all types.

%% file: s9_conclusion.tex
\section{Conclusion}
\label{sec:conclusion}

We presented a novel visual metaphor for the visualization of hierarchically structured data over time.
Our approach allows for the clear representation of hierarchies at certain points in time, while simultaneously conveying the temporal evolution of data values and changes in the hierarchy. 
Compared to existing techniques, all possible hierarchical changes are supported and represented. 
The evaluation confirms that our approach provides equivalent performance to treemaps and nested streamgraphs in the analyzed tasks they perform best in and therefore makes a good general-purpose technique.
We provide a JavaScript library for the easy reproduction of all demonstrated examples and for integration into other projects at \textit{https://github.com/cadanox/SplitStreams}.